\documentclass[final,3p,times,number]{elsarticle}

\usepackage{amssymb}
\usepackage{amsmath}
\usepackage{amsthm}
\usepackage{graphicx}
\usepackage{booktabs}
\usepackage{enumitem}
\usepackage{mathtools}
\usepackage{subcaption}
\usepackage{caption}
\usepackage{float}
\usepackage{nomencl}
\usepackage{bm}
\usepackage{tabularx}
\usepackage{xurl}
\usepackage[hidelinks]{hyperref}
\usepackage[capitalize]{cleveref}
\journal{Energy Conversion and Management}

\begin{document}

\begin{frontmatter}



\title{Quantifying the power grid hosting capacity gained by coordinating distributed solar, battery storage and electrified flexible loads} 

\author[a]{Vineet Jagadeesan Nair\corref{cor1}}
\ead{jvineet9@mit.edu}
\author[b]{Morteza Vahid Ghavidel}
\author[a]{Anuradha M. Annaswamy\corref{cor1}}
\ead{aanna@mit.edu}
\cortext[cor1]{Corresponding authors}
\affiliation[a]{organization={Department of Mechanical Engineering, Massachusetts Institute of Technology},
            addressline={77 Massachusetts Avenue},
            city={Cambridge},
            postcode={02139},
            state={MA},
            country={USA}}

\affiliation[b]{organization={School of Technology and Innovations, University of Vaasa},
            addressline={Wolffintie 32},
            city={Vaasa},
            postcode={65200},
            country={Finland}}

\begin{abstract}
Distribution grid hosting capacity is usually assessed with static, single-resource methods that treat distributed energy resources as fixed injections, which can understate how much solar generation and electrified load feeders can accept. We quantify the hosting capacity gained when solar photovoltaics (PV), battery storage, heat pumps and electric vehicles (EVs) are dispatched by network-aware coordination at 5-minute resolution, versus static operation of the same devices. On the IEEE 123-node feeder, we use a deterministic iterative method, a two-stage stochastic program that optimizes siting, and a Monte Carlo extension over 100 scenarios. With storage, heat pumps and EVs at 5\%, coordination raises PV hosting capacity from 48\% to 83\% of peak load; with 10\% storage, 20\% PV and 5\% EVs, heat pump hosting capacity rises from 9\% to 55\% of homes. When siting is co-optimized, PV hosting capacity rises from 24\% to 98\%, and the jointly feasible PV, heat pump and storage region grows about 22-fold. Coordination turns storage and heat pumps from substitutes into complements: with storage up to 100\% of peak load, the feeder hosts PV up to about 200\% of peak load and heat pumps in 90\% of homes. Midday limits are relieved by storage charging and heat pump pre-cooling, evening limits by inverter reactive support, battery discharging and heat pump setback. Under input uncertainty, coordination raises mean PV hosting capacity by 64\%; the absolute spread widens while the relative spread narrows slightly. At published reinforcement costs, the added capacity is worth roughly \$0.6--1.5 million per feeder in deferred upgrades.
\end{abstract}


\begin{highlights}
\item Coordinating DERs raises PV hosting capacity from 48\% to 83\% of peak load
\item Heat pump hosting capacity rises from 9\% to 55\% of homes under coordination
\item Coordination turns storage and heat pumps from substitutes into complements
\item The jointly feasible PV, heat pump and storage region grows about 22-fold
\item Coordination raises mean hosting capacity by 64\% but widens its absolute spread
\end{highlights}


\end{frontmatter}

\noindent\fbox{\begin{minipage}{0.97\textwidth}
\footnotesize
\textbf{Nomenclature}\\[2pt]
\begin{minipage}[t]{0.48\textwidth}
\textit{Abbreviations}\\[1pt]
\begin{tabular}{@{}p{1.25cm}p{5.35cm}@{}}
2-SSP & two-stage stochastic program \\
ACOPF & AC optimal power flow \\
BS & battery storage \\
DER & distributed energy resource \\
DERMS & DER management system \\
DI & deterministic iterative (method) \\
DSO & distribution system operator \\
EV & electric vehicle \\
HC & hosting capacity \\
HCA & hosting capacity analysis \\
HP & heat pump \\
LMP & locational marginal price \\
MISOCP & mixed-integer second-order cone program \\
PCC & point of common coupling \\
PV & solar photovoltaics \\
SAA & sample average approximation \\
SI & stochastic iterative (method) \\
SOC & state of charge \\
V2G & vehicle-to-grid \\
\end{tabular}\\[3pt]
\textit{Sets and indices}\\[1pt]
\begin{tabular}{@{}p{1.25cm}p{5.35cm}@{}}
$\mathcal{N}, \mathcal{E}$ & nodes and branches of the feeder \\
$i, j$ & node indices; $(i,j)$ a branch \\
$t, \Delta t$ & time step and its length (5 min) \\
$\mathcal{T}, \mathcal{H}$ & simulation period and planning horizon \\
$\Omega, \omega_k$ & scenario set and scenario $k$ \\
$K, p_k$ & number of scenarios; probability of $\omega_k$ \\
\end{tabular}\\[3pt]
\textit{Network}\\[1pt]
\begin{tabular}{@{}p{1.25cm}p{5.35cm}@{}}
$P_i, Q_i$ & net active and reactive injection at node $i$ \\
$P_{ij}, Q_{ij}$ & active and reactive branch flow \\
$v_i, l_{ij}$ & squared voltage and squared current magnitude \\
$R_{ij}, X_{ij}$ & branch resistance and reactance \\
$\overline{S}_{ij}, S_{tran}$ & branch and transformer apparent power limits \\
$P_i^L, Q_i^L$ & fixed (inflexible) load \\
$\underline{pf}$ & minimum inverter power factor \\
\end{tabular}
\end{minipage}\hfill
\begin{minipage}[t]{0.48\textwidth}
\textit{DER models}\\[1pt]
\begin{tabular}{@{}p{1.8cm}p{4.8cm}@{}}
$P_i^{PV}, \overline{P}_i^{PV}$ & PV output and rated capacity \\
$\alpha^{PV}(t)$ & normalized solar irradiance profile \\
$P_i^{BS}, \overline{P}_i^{BS}$ & BS power (positive when discharging) and rating \\
$\overline{E}_i^{BS}, E_r^{BS}$ & BS energy capacity; energy-to-power ratio \\
$SOC_i^{BS}$, $SOC_i^{EV}$ & state of charge of BS and EV \\
$\eta^{BS}_i, \delta^{BS}_i$ & BS efficiency and self-discharge rate \\
$P_i^{EV}$ & EV charger power (positive when discharging) \\
$P_r^{EV}, E_r^{EV}$ & EV charger rating and battery capacity \\
$P_i^{HP}, P_r^{HP}$ & HP electrical power and unit rating \\
$T_i^{in}, T^{out}$ & indoor and outdoor temperature \\
$T_i^{in^*}$ & indoor temperature setpoint \\
$R^{th}_i, C^{th}_i, \eta_i$ & thermal resistance, capacitance and COP \\
$\bar{\Delta}_{hp}$ & HP ramp limit (fraction of $P_r^{HP}$) \\
\end{tabular}\\[3pt]
\textit{Planning and optimization}\\[1pt]
\begin{tabular}{@{}p{1.8cm}p{4.8cm}@{}}
$x_i^{PV}, x_i^{BS}$ & PV and BS penetration at node $i$ (fraction of nodal peak load) \\
$x_i^{EV}, x_i^{HP}$ & share of homes at node $i$ with an EV or HP \\
$\overline{x}^{BS}$ & feeder storage penetration set by the planner \\
$\overline{P^L}_{nom,i}$ & nominal peak load at node $i$ \\
$n_i^h$ & number of homes at node $i$ \\
$f^{I}, f^{II}$ & first- and second-stage objectives \\
$f^{OPF}$ & coordinated dispatch objective \\
$\beta$ & penalty weights in $f^{OPF}$ \\
$\lambda(t), \bar{\lambda}$ & real-time LMP and its mean \\
\end{tabular}
\end{minipage}
\end{minipage}}
\medskip

\section{Introduction}
\label{sec:intro}

In order to decarbonize the grid, we need to rapidly integrate renewables like wind and solar, in addition to battery storage. Furthermore, the electrification of other residential sectors like heating and transportation is causing rapid growth in electricity demand. We expect to see an additional 262 GW of distributed energy resources (DERs) in the distribution grid between 2023 and 2027, which is comparable to the growth in utility-scale resources (+272 GW) at the transmission grid level over the same period \cite{woodmac2023der}, and corresponds to billions of dollars in new investments. The current conversation tends to be dominated by concerns around transmission grid capacity and utility-scale interconnection, particularly with data center expansion. However, it is important to ensure that the distribution grid also has enough capacity to accommodate the growth in installations of such small-scale resources. This paper explores how a distributed coordination and optimization approach can help the distribution grid reliably and affordably accommodate more DERs. We assume that all the DERs are autonomous and independently owned. Coordination of such resources in real time can be achieved through new retail market structures, price signals or other incentive programs offered by utilities and load-serving entities. Certain DERs may also be directly operated by utilities or grid operators. Regardless of the specific implementation mechanism involved, our optimization approach and framework provide stakeholders with decision-making tools to optimally coordinate and manage their assets to enhance hosting capacity and grid reliability. In turn, coordination can mitigate the voltage and thermal loading issues that would otherwise limit how much new solar generation and electrified load a feeder can accept. Viewed as an energy-management strategy, coordination converts existing feeder capacity into usable capacity for energy conversion and storage devices, including PV inverters, heat pumps, batteries and EV chargers.

Hosting capacity (HC) refers to the maximum amount of DERs such as solar PV, battery storage, electric vehicles (EVs), and heat pumps (HPs) that can be integrated into the electric grid at a given location without causing reliability, power quality, or safety violations and without requiring major system retrofits or upgrades. It is typically expressed in absolute terms or as a percentage of the peak load at that location. This is a critical metric for utilities and grid operators to assess the feasibility of integrating renewable energy sources and other distributed generation technologies, as well as new loads into the existing distribution grid. In order to determine the HC for additional DERs, we aggregate constraints related to voltage, thermal, or protection limits over the whole distribution feeder and use them for high-level planning \cite{Seuss2015ImprovingSupport}.

HC may be limited by a variety of voltage, power quality, reliability, thermal, or operational constraints. In this work, we focus on nodal voltage magnitude limits and line thermal (current) constraints, which are the two binding constraint categories most commonly reported in distribution-level HC studies \cite{Mousa2024ANetworks,Koirala2022HostingApproaches}. Overvoltage and undervoltage limits are the primary voltage concerns: excess solar PV output during the middle of the day drives overvoltage, while heavy electrified loads at the end of feeders cause undervoltage. Apart from voltages, thermal limits on maximum line currents (equivalently, maximum apparent power flows) are the other major binding factor. We also include the substation transformer's rated capacity as an additional constraint. Other power quality metrics such as harmonic distortion and flicker are relevant for individual device interconnection studies but are generally not the binding constraints at the feeder planning level addressed here; we therefore treat them as out of scope and a direction for future work.

Regarding voltage imbalances, these can arise in unbalanced three-phase systems with asymmetric DER placement. In this study, we assume either a balanced system or convert the network to a balanced equivalent, which is a standard simplification in medium-voltage distribution HC analysis \cite{Molzahn2017ASystems}. Extensions to explicitly unbalanced multi-phase networks are a natural direction for future work. In terms of DER scope, we focus on solar PV, battery storage, EVs, and heat pumps as these represent the four largest and fastest-growing categories of distributed resources in residential and commercial distribution feeders today \cite{woodmac2023der}. Other technologies such as small wind turbines and micro-CHP units have comparatively lower projected penetration in the low-to-medium voltage distribution networks that are the focus of this study, and their inclusion is left for future work.

Many utilities across the US are already severely HC-constrained, which delays or prevents the integration of new DERs in the distribution grid. The issue of hosting capacity is also closely related to the problem of renewables curtailment, wherein excess solar PV output in the middle of the day is increasingly being diverted or wasted simply because the grid does not have sufficient capacity to transport that power. This leads to overbuilding of capacity, increased capital costs, and higher electricity rates. Ideally, we would like to avoid curtailment of clean power and instead utilize PV output productively (e.g., to charge batteries or EVs). This is another motivation behind this work. Throughout, we evaluate hosting capacity without any PV curtailment, the stricter definition, so that any capacity gained through coordination comes from using PV output rather than discarding it.

\section{Literature review}
\label{sec:lit}

Hosting capacity analysis (HCA) methods have evolved considerably over the past decade, progressing from simple worst-case estimates to scenario-based frameworks. Existing methodologies can be broadly categorized into deterministic, stochastic/probabilistic, and time-series approaches \cite{Mousa2024ANetworks,Koirala2022HostingApproaches}. In static or deterministic HCA, the distribution system is evaluated under fixed extreme conditions such as maximum PV output coinciding with minimum load, yielding conservative HC values that are computationally tractable but may not reflect operational reality \cite{Fatima2020ReviewNetworks}. While these approaches provide a useful baseline, they fail to capture the variability inherent in renewable generation and demand, which can lead to either underestimation or overestimation of true hosting limits and potentially misguide investment and interconnection decisions. Recent reviews emphasize that both steady-state and transient constraints should be included in HC studies, and that reliance on a single worst-case snapshot is increasingly untenable as DER penetration grows \cite{Xie2021TowardReady,Mousa2024ANetworks}.

Time-series (quasi-static) methods extend static analyses by simulating the feeder over hours, days, or seasons at high temporal resolution, capturing diurnal and seasonal variability in load and generation \cite{Makhalfih2023DistributedPerspective,TavaresdeOliveira2025APenetration}. For example, \cite{Makhalfih2023DistributedPerspective} and \cite{Mulenga2021SolarUncertainties} evaluate hosting capacity at critical hours, i.e., midday solar peaks and evening load peaks, and generally find higher capacity than static worst-case analyses, since constraints such as overvoltage bind only intermittently. To simplify the analysis, some studies adopt a worst-case time-period approach and consider only specific hours with peak PV output or peak load \cite{Mousa2024ANetworks}. Tavares de Oliveira et al.\ \cite{TavaresdeOliveira2025APenetration} further demonstrate that incorporating time-varying medium-voltage background voltages with increasing penetration reduces the estimated hosting capacity by 32\% on average, underscoring the importance of multi-day or year-long horizons to fully capture diverse operating scenarios. Moreover, these methods can be computationally intensive and, on their own, do not incorporate uncertainty in a principled stochastic framework.

Stochastic and probabilistic methods address these limitations by explicitly modeling uncertainties in DER output, load profiles, and network parameters. Approaches include Monte Carlo simulation, probability distribution functions, and chance-constrained optimization, which yield a distribution of hosting capacity outcomes and can quantify the probability of constraint violations to enable risk-informed planning \cite{Geng2021ProbabilisticOptimization,Koirala2022HostingApproaches}. The authors in \cite{Koirala2022HostingApproaches} conduct a systematic benchmark of deterministic and stochastic hosting capacity definitions on a real semi-urban network, proposing a reference stochastic method that considers all sources of uncertainty as well as operational and probabilistic limits. Moreover, the authors in \cite{Geng2021ProbabilisticOptimization} cast probabilistic HCA as a chance-constrained optimization solved via Bayesian optimization, demonstrating a 25\% improvement in hosting capacity with substantially less computation than traditional solvers. Mulenga et al.\ \cite{Mulenga2021SolarUncertainties} further distinguish between aleatory (inherent randomness) and epistemic (knowledge-based) uncertainties in solar PV hosting capacity, using superposition and reduced time-domain sampling to efficiently capture both. The authors in \cite{Taheri2021FastSystems} leverage multiparametric programming to accelerate probabilistic HCA, solving over 500,000 optimal power flow instances by computing only a small fraction of them. More recently, Lee et al.\ \cite{Lee2024AnSystems} introduce active learning into HCA to efficiently explore the vast space of DER integration scenarios, showing that hosting capacity and its interpretations change significantly depending on DER adoption patterns and control strategies. In general, probabilistic HCA provides more nuanced insight such as the probability of voltage violations, and higher capacity estimates by relaxing worst-case assumptions, but at the cost of much larger scenario sets. Critically, most probabilistic studies evaluate HC for given DER placements and operating rules rather than choosing the placements and dispatch themselves, and few integrate coordination mechanisms or price-based incentives into the analysis.

A complementary line of work embeds control and planning decisions directly into the HCA, actively enhancing capacity rather than merely evaluating it. For instance, \cite{Abad2021PhotovoltaicManagement} proposes a framework based on optimization to determine PV hosting capacity considering autonomous voltage control strategies including active power curtailment, reactive power control, and on-load tap changer management, and demonstrates their effectiveness across 128 low-voltage UK feeders. Moreover, the authors in \cite{Yuan2022DeterminingGrids} propose a constructive geometric model for HC determination that preserves global optimality even with non-convex formulations, achieving larger hosting capacity values with less computation time than prior approaches. Wu et al.\ \cite{Wu2022RobustCorrelation} develop a robust comprehensive PV hosting capacity assessment model that considers both spatial and temporal correlations of PV output using ellipsoidal uncertainty sets, showing that considering these correlations significantly increases the estimated hosting capacity. A hierarchical bi-level model in which the lower level optimizes customer demand via demand response and dynamic tariffs, while the upper level simultaneously maximizes DG and EV capacity, achieving dramatic increases in allowable DER penetration on both the IEEE 33-bus test system and a real 59-bus Cairo network, is proposed in \cite{Zenhom2025SimultaneousControl}. The authors in \cite{Gkontoras2025Bi-levelUpgrades} develop a bi-level framework that optimizes both capital and operational costs of grid expansion plans considering smart grid technologies and conventional upgrades, demonstrating that leveraging load-shifting flexibilities can defer costly asset replacements. These works collectively demonstrate that incorporating active network management such as reactive power support, topology control, and flexible load into HCA can substantially expand capacity relative to passive scenarios. However, most optimization-based HC studies consider only a single DER type (typically solar PV) and do not jointly optimize multiple DER categories within a unified framework.

Another line of research examines DER coordination. Whereas most prior HCA assumes autonomous DER operation, emerging studies quantify the benefits of coordinated scheduling. For example, \cite{Navidi2023CoordinatingLoad} uses a power-flow-driven simulation and optimization framework to compare default local DER control against a perfect-foresight centralized controller and finds that central coordination sharply reduces grid violations; for example, the fraction of overloaded transformers falls from approximately 81\% under local control to 28\% under centralized control by 2050, while peak load is reduced by 17\%. Moreover, a hybrid volt-var control architecture leveraging distributed optimization for DER coordination in active distribution grids is demonstrated in \cite{Haider2022AGrids}, showing that even in networks with existing voltage control devices, explicit DER coordination is necessary to maintain acceptable voltage profiles. Hwang et al.\ \cite{Hwang2023DemandStudy} present hosting capacity enhancement methods based on optimal price-based demand response of heating, ventilation and air conditioning (HVAC) systems in commercial buildings, using a bi-level decision model that considers both distribution system operator (DSO) profits and end-user costs. Recent advances also leverage data-driven approaches including machine learning, deep reinforcement learning, and physics-informed learning for real-time, adaptive DER control and decision-making \cite{Lee2024AnSystems,Yu2025Data-drivenNetworks}. However, most coordination studies focus on operational control rather than joint planning and design, and few explicitly quantify the resulting improvement in hosting capacity within an optimization framework.

Finally, a growing body of work recognizes that hosting capacity is not only a technical limit but also a socio-economic and regulatory challenge. For instance, \cite{Brockway2021InequitableCalifornia} highlights that existing grid constraints lead to inequitable DER access, with low-income and disadvantaged communities often facing much lower hosting capacity on their circuits. Gorman et al.\ \cite{Gorman2025} document severe grid interconnection bottlenecks at the bulk level: U.S.\ queues now exceed system capacity, with new requests facing long delays and high withdrawal rates. Xie et al.\ \cite{Xie2021TowardReady} warn that distribution equipment and protection schemes, designed for one-way power flows, will increasingly be violated under high DER penetration, even with flat demand profiles. Meanwhile, it is estimated that accommodating full electrification of housing and personal vehicles across the U.S.\ could require up to 312~GW of distribution grid reinforcement at a cost of \$183--\$415 billion in \cite{Priyadarshan2024DistributionElectrification}, though demand-side management can reduce these costs by up to 92\%. Moreover, the authors in \cite{Aydin2025} further demonstrate that fairness-aware dynamic hosting capacity approaches with strategic solar PV curtailment of no more than 5\% can increase hosting capacity by at least 50\% with net positive economic impact. These findings underscore that maximizing hosting capacity also requires addressing regulatory, planning, and equity dimensions --- a context in which technically rigorous, optimization-driven HCA methods become important.

\cref{tab:lit_comparison} compares representative studies along the dimensions that matter for this paper: which DER types are hosted, whether DERs are actively managed or coordinated, how uncertainty is treated, whether siting and sizing are optimized, and whether the gain over a static baseline is reported as hosting capacity.

\begin{table}[htbp]
    \centering
    \footnotesize
    \begin{tabularx}{\textwidth}{@{}>{\raggedright\arraybackslash}p{2.3cm}>{\raggedright\arraybackslash}p{1.7cm}>{\raggedright\arraybackslash}X>{\raggedright\arraybackslash}p{2.4cm}>{\raggedright\arraybackslash}p{1.2cm}>{\raggedright\arraybackslash}p{2.6cm}@{}}
    \toprule
    Study & DER types & Active management or coordination & Uncertainty & Siting/ sizing & Outcome reported \\
    \midrule
    Seuss et al.\ \cite{Seuss2015ImprovingSupport} & PV & Local inverter volt-var control & -- & No & PV HC with and without reactive support \\
    Torquato et al.\ \cite{Torquato2018ASystems} & PV & None & Monte Carlo & No & PV HC of LV networks \\
    Koirala et al.\ \cite{Koirala2022HostingApproaches} & PV & None & Deterministic vs.\ stochastic benchmark & No & PV HC under alternative definitions \\
    Abad et al.\ \cite{Abad2021PhotovoltaicManagement} & PV & Curtailment, reactive power, OLTC & Probabilistic & No & PV HC sensitivity, 128 LV feeders \\
    Taheri et al.\ \cite{Taheri2021FastSystems} & DER (generic) & OPF-based DER setpoints, voltage regulators & Probabilistic (multiparametric) & No & Fast probabilistic HC \\
    Geng et al.\ \cite{Geng2021ProbabilisticOptimization} & DER (generic) & -- & Chance constraints & No & Probabilistic HC \\
    Wu et al.\ \cite{Wu2022RobustCorrelation} & PV & Active network management & Robust (spatio-temporal correlation) & No & PV HC \\
    Zhan et al.\ \cite{Zhan2022TowardsCapacity} & HP & Congestion management & -- & No & HP HC \\
    Hwang et al.\ \cite{Hwang2023DemandStudy} & HVAC & Price-based demand response (bi-level) & -- & No & HC improvement from HVAC flexibility \\
    Zenhom et al.\ \cite{Zenhom2025SimultaneousControl} & PV, wind, EV & Demand response, dynamic tariffs, volt/VAR (bi-level) & -- & Sizing & Joint DG and EV HC \\
    Navidi et al.\ \cite{Navidi2023CoordinatingLoad} & PV, BS, EV, electrified load & Centralized vs.\ local coordination & Perfect foresight & No & Violations, peak load, upgrades (not HC) \\
    \textbf{This work} & PV, BS, EV, HP & Network-aware receding-horizon dispatch, 5-min & Deterministic; two-stage stochastic; Monte Carlo (100) & Yes (2-SSP) & PV and HP HC gain over static baseline; joint envelope; deferral value \\
    \bottomrule
    \end{tabularx}
    \caption{Comparison with representative hosting capacity and DER coordination studies. ``--'' indicates that the dimension is not a focus of the study.}
    \label{tab:lit_comparison}
\end{table}

\subsection{Research gap and contributions}

The majority of prior HC works have focused on maximizing only solar PV HC \cite{Seuss2015ImprovingSupport}. Some works have assessed the HC of either EV, HVAC, or HPs \cite{Zhan2022TowardsCapacity}, but generally in isolation. Jointly analyzing multiple DER types introduces three challenges that are absent from single-DER studies: (i) the feasible hosting capacity of each DER type is not independent --- it depends on the penetrations and dispatch of all other types simultaneously; (ii) the complementary or competitive relationships among DER types are not known a priori and must be discovered through optimization; and (iii) the joint planning problem requires determining both the siting and sizing of multiple DER types across all network nodes, which grows combinatorially with the number of types and nodes. These challenges mean that running separate single-DER analyses and aggregating the results can misestimate the joint feasibility space.

Coordination studies report fewer constraint violations \cite{Navidi2023CoordinatingLoad,Haider2022AGrids}, and HC studies report limits for one or two DER types \cite{Seuss2015ImprovingSupport,Zhan2022TowardsCapacity,Zenhom2025SimultaneousControl}, but none measures how much hosting capacity coordinating PV, storage, heat pumps and EVs adds over a consistent static baseline, which mechanism supplies that capacity, and how it behaves under uncertainty. This paper addresses that gap. Its contributions are the following.
\begin{enumerate}
    \item \textit{The coordination dividend.} Against a static baseline that operates the same devices without coordination, network-aware coordination raises PV HC from 48\% to 83\% of peak load (a 73\% relative increase) with storage, heat pumps and EVs at 5\%, and heat pump HC from 9\% to 55\% of homes with 10\% storage, 20\% PV and 5\% EVs. When DER siting and sizing are also co-optimized, PV HC rises from 24\% to 98\%.
    \item \textit{The joint hosting envelope.} We extend HC from a scalar limit to a feasible penetration envelope for PV, heat pumps and storage, computed with a two-stage mixed-integer second-order cone stochastic program that co-optimizes DER siting and sizing. Coordination enlarges this envelope about 22-fold and turns storage and heat pumps from substitutes into complements.
    \item \textit{Mechanism attribution.} By examining the constraints during the hours when they bind, we show which resource relieves which limit: midday overvoltage and line loading from reverse flow are relieved by storage charging and heat pump pre-cooling; evening undervoltage by inverter reactive support, battery discharging and heat pump setback; and line loading mainly by heat pump flexibility.
    \item \textit{Behaviour under uncertainty.} Over 100 sampled input scenarios, coordination raises mean PV HC by 64\%; the absolute spread widens while the relative spread narrows slightly, so planners should size coordinated HC to a low percentile of its distribution rather than to its mean.
    \item \textit{Decision value.} We translate the added capacity into deferred reinforcement value per feeder, set it against the break-even cost of a coordination layer, and map each method to an existing utility planning or operating process.
\end{enumerate}
Each method embeds the device-level dynamics of all four DER types, including intertemporal state constraints for batteries, EVs and heat pumps, and we keep the stochastic program tractable with k-means scenario reduction, Monte Carlo sampling and warm starting. These are the tools that produce the results above rather than contributions in their own right.

The remainder of the paper is organized as follows. \cref{sec:methods} describes the input data, the network and DER models, and the three HC methods. \cref{sec:results} reports the results of each method and reconciles them. \cref{sec:discussion} discusses deferral value, deployment pathways and limitations, and \cref{sec:conclusion} concludes.

\section{Methods}
\label{sec:methods}

In this work, we propose and compare three complementary HC analysis approaches, each designed for a distinct decision-making context. The \textit{deterministic iterative} (DI) method is intended for system operators and engineers who need transparent, interpretable insight into how DER interactions and dynamic dispatch affect grid metrics at specific penetration levels. Each step requires only one power flow or optimal power flow solve, which makes the method simple to implement and interpret and well-suited for sensitivity analysis and scenario exploration, although a full sweep over penetration levels requires many such solves. The \textit{two-stage stochastic program} (2-SSP) targets grid planners and utilities who must make longer-horizon investment and design decisions --- it jointly optimizes DER siting and sizing across the network while enforcing feasibility in every scenario of a representative set of operating scenarios. Finally, the \textit{stochastic iterative} (SI) method, which embeds Monte Carlo sampling within the DI framework, is designed for risk-aware planning: it quantifies the full distribution of HC outcomes under input uncertainty, providing planners with conservative estimates and confidence intervals. Using all three methods allows us to cross-validate results and capture aspects of the HC problem that no single approach addresses alone.

\subsection{Input data}

The key inputs into our simulation are data related to DER power injections for both generation and load. For the demand side, this includes load profiles and temperature data, which influence HP use. On the supply side, solar irradiation data affect the PV output. In addition to the baseline injection profiles, there are also distributions of parameters that influence the amount of flexibility (along with its temporal signature) that customers (or prosumers) are willing to provide. These include both downward and upward flexibilities in generation and load. For the static cases without flexibility, we derived several representative scenarios and profiles for DER power injections from real historical data and smart meter measurements. These specify how prosumers operate their DERs in the nominal case \cite{pecanstreet_dataport}. It is especially important to have accurate estimates of net injection profiles for BS, EV, and HP operation, while PV output will follow irradiance profiles. We use real-time locational marginal price (LMP) data from the California Independent System Operator (CAISO) \cite{CAISO_OASIS}, which influences how the DSO and agents balance importing power from the main transmission grid versus relying on the dispatch of local assets like batteries and EVs. The remaining parameters are generated randomly for each feeder node and assumed to be the same for all scenarios of inputs. For example, distributions of the key EV parameters can be obtained from actual load profiles of EV charging data. These include the unavailable time windows and state of charge targets for each EV. \cref{tab:input_data} summarizes each input, its source and resolution, and how it enters the scenarios.

Our simulations are conducted using a version of the IEEE 123-node test feeder \cite{schneider2017analytic}. This benchmark has become a standard reference for medium-voltage distribution HC studies because it captures key features of realistic US feeders: a heterogeneous mix of residential, commercial, and industrial loads; uneven spatial load distribution; multiple laterals of varying length; and a range of impedances that create realistic voltage gradients. We modify it to support high penetrations of renewables, storage, and flexible loads following the approach of \cite{Navidi2023CoordinatingLoad}, including adding distributed PV, battery, EV, and HP assets at nodes proportional to load size. While no single test feeder can be universally representative, the 123-node system is widely used as a proxy for North American medium-voltage distribution feeders and its results are broadly indicative of similarly structured radial networks \cite{Mousa2024ANetworks}. Extensions to low-voltage (LV) secondary networks --- which introduce additional complexity from phase imbalance and low X/R ratios --- are an important direction for future work. Our approach can be applied to other radial distribution networks, which represent the majority of utility feeders in the United States \cite{Navidi2023CoordinatingLoad}.

\begin{table}[htbp]
    \centering
    \small
    \begin{tabularx}{\textwidth}{@{}p{3.0cm}Xp{1.5cm}X@{}}
    \toprule
    Input & Source & Resolution & Use in scenarios \\
    \midrule
    Fixed (inflexible) load $P_i^L(t)$ & Smart-meter measurements \cite{pecanstreet_dataport} & 5 min & Baseline profile plus Gaussian noise; differs by node and scenario \\
    Static HP and EV profiles & Profiles derived from historical measurements \cite{pecanstreet_dataport} & 5 min & Inflexible injections in the static case; perturbed by Gaussian noise \\
    Static BS profile & Local self-consumption rule within SOC limits (\cref{sec:coordination}) & 5 min & Inflexible injection in the static case \\
    Solar irradiance $\alpha^{PV}(t)$ & Representative clear summer day, San Francisco, CA; 5-min irradiance for this location is published in the NSRDB \cite{Sengupta2018NSRDB,NSRDB_GOES_v4} & 5 min & Normalized; common to all nodes; perturbed by Gaussian noise \\
    Outdoor temperature $T^{out}(t)$ & Air temperature for the same representative day and location; see NSRDB \cite{Sengupta2018NSRDB,NSRDB_GOES_v4} & 5 min & Common to all nodes; perturbed by Gaussian noise \\
    Real-time LMP $\lambda(t)$ & CAISO OASIS \cite{CAISO_OASIS} & 5 min & Common to all nodes; perturbed by Gaussian noise \\
    EV availability windows and SOC targets & EV charging data \cite{CastilloJr.2022UsingCommercialization} & Per node & Sampled per node; fixed across scenarios \\
    HP thermal parameters & \cite{Hao2015AggregateLoads}; \cref{tab:hp-parameters} & Per home & Fixed \\
    Network & IEEE 123-node feeder, 2,930~kW peak load \cite{schneider2017analytic,IEEE123BusTestSystem} & -- & Fixed topology, impedances and nodal loads \\
    \bottomrule
    \end{tabularx}
    \caption{Input data. Scenarios are built by adding independent Gaussian noise to each baseline profile. The stochastic iterative method uses 100 such scenarios; the 2-SSP uses 5 representative scenarios selected from them by k-means clustering (\cref{sec:accel_2ssp}).}
    \label{tab:input_data}
\end{table}

\subsubsection{Network configuration and defining DER penetration}

DER penetration is defined slightly differently depending on the type of DER. PV and BS penetrations are defined relative to the peak feeder load, with no upper limit or cap. HP and EV penetrations are defined in terms of the proportion (\%) of homes/buildings in the feeder that have been electrified, and the upper limit is 100\%. All air conditioners for cooling are already electric. But heating (both space and water) is largely still gas-based and will transition to electric heat pumps. These heat pumps (either retrofitted or newly installed) can also act as cooling devices, and the average heat pump (HP) load for a home is about 5-6 kW. The average US house load is 1.2 kW with a peak load of 3-6 kW \cite{EIA_Residential_2024}. The base load (fixed) is 0.5-0.8 kW (from appliances like refrigerators, standby electronics, etc.), while the peak load is from thermostatically controlled loads (TCLs) like ACs, heat pumps, water heaters, electric dryers, etc. In our IEEE 123-node feeder configuration, the total nominal peak load of the feeder is 2,930~kW. We estimate the number of homes at each node as its nominal peak load divided by an average residential peak load of 3.5~kW, which gives approximately 840 homes on the feeder. The box below gives a worked example of each definition.

\medskip
\noindent\fbox{\begin{minipage}{0.96\textwidth}
\textbf{How DER penetration and hosting capacity are defined.}\\[3pt]
\textit{PV and storage} are measured relative to the feeder's nominal peak load of 2,930~kW, with no upper limit. A PV penetration of 48\% means about $0.48 \times 2{,}930 \approx 1{,}410$~kW of PV rated capacity installed across the feeder; 200\% means PV capacity twice the peak load. A storage penetration of 5\% means 147~kW of battery power with 441~kWh of energy (3-hour duration).\\[3pt]
\textit{Heat pumps and EVs} are measured as the share of the feeder's approximately 840 homes that have one, capped at 100\%. A heat pump penetration of 9\% means about 75 homes have a heat pump; 55\% means about 460 homes.\\[3pt]
\textit{Hosting capacity} of a DER type is the highest penetration of that type at which nodal voltages, line currents and the substation limit stay within bounds at every simulated time step (and, for the 2-SSP, in every scenario), with no PV curtailment, for given penetrations of the other DER types. Storage penetration is always an input, never a hosting capacity.
\end{minipage}}
\medskip

\subsection{Increasing hosting capacity with coordination}
\label{sec:coordination}

\begin{figure}
    \centering
    \includegraphics[width=\textwidth]{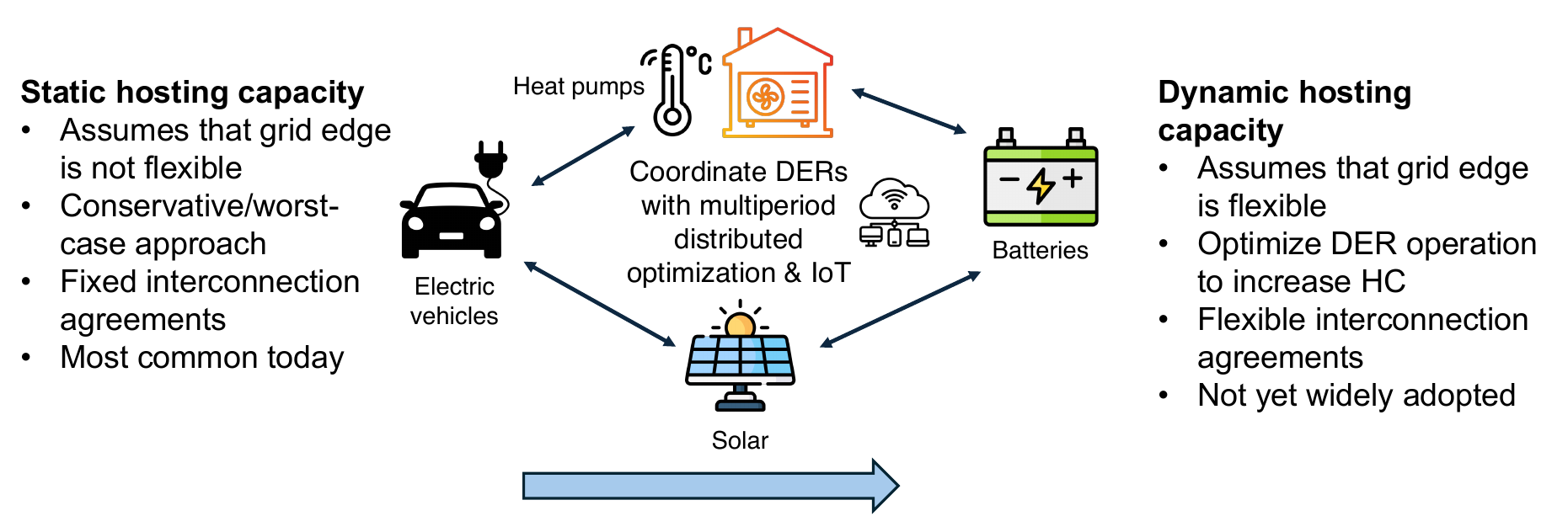}
    \caption{Approach for increasing hosting capacity: dynamic coordination exploits the flexibility and complementarity of generation, storage and flexible demand, in contrast to static analysis with fixed DER injections.}
    \label{fig:hc_increase_overview}
\end{figure}

\cref{fig:hc_increase_overview} provides an overview of the approach we use to increase the hosting capacity. We leverage complementary relationships among different types of DERs (distributed generation, storage, flexible demand) and harness the flexibility available at the grid edge to dynamically increase hosting capacity. This coordination can aid both grid planning and operation, and requires flexible interconnection agreements, which are not yet widely adopted by utilities. This is in contrast to the static HC analysis, which takes a conservative or worst-case approach based on fixed interconnection agreements.
In what follows, we show that dynamic coordination and receding-horizon optimization, executed at 5-minute intervals that match real-time market settlement, can increase grid hosting capacity and reduce voltage and line-loading stress. We use the term ``real-time'' throughout to refer to this sub-hourly dispatch timescale, as distinct from slower day-ahead or seasonal planning horizons.

For the static case, we use inflexible baseline profiles for fixed loads, heat pumps, batteries, and solar radiation profiles for PV. We derive each of these profiles from historical data and measurements. Considering many scenarios also allows us to move towards non-deterministic or probabilistic results as well. We assume that these injections are not flexible and that there is no coordination or optimization among the DERs. We simply run the power flow model and check if the DER power injections are AC-feasible. The heat pump and electric vehicle baseline profiles are based on real-world data, while solar radiation profiles correspond to a location near San Francisco, CA. For the static battery charging or discharging profiles, we assume that the batteries are uncoordinated across nodes --- batteries at each node are operated to maximize self-consumption while respecting state of charge (SOC) limits, but in a localized and myopic manner without considering the rest of the network or planning for future time periods.

For the dynamic case, we leverage network-wide global coordination over the whole feeder. The dynamic approach consists of the following steps:
\begin{enumerate}
    \item \textbf{Coordination} of different types of DERs at each node. This determines the flexibility in power injections available, while respecting all device-level operational constraints and user preferences.
    \item Solve the network-wide \textbf{multiperiod} optimization problem.
    \item Communicate the optimal solution through the chosen \textbf{coordination mechanism}, e.g., a distributed energy resource management system (DERMS), operating envelopes, or a distribution-level market.
    \item Schedule or operate the DERs based on their updated \textbf{dispatch setpoints} (e.g., power output, temperature).
\end{enumerate}

This allows collaboration among different nodes rather than each agent just optimizing its own DERs locally. This is enabled by solving the global AC optimal power flow problem. This global optimization can be solved either in a centralized manner or by using distributed algorithms which are privacy-preserving and more computationally efficient \cite{Romvary2022AOptimization}. A distribution-level retail market structure could be one approach to achieve this coordination, which would also allow us to compensate resources for their flexibility through pricing. However, we can also do this without markets, as long as there is infrastructure in place for communication and coordination among the different operators and agents.

\subsection{AC power flow model}

We use the branch flow model with second-order conic program (SOCP) convex relaxation \cite{Molzahn2017ASystems,Molzahn2019AEquations}, where $P_i^L$ refers to fixed loads. Under mild assumptions, this relaxation is exact for radial (tree) grids, which are the most common topology for distribution networks. If the system is multi-phase and unbalanced, we convert it to an equivalent balanced system. Extensions to meshed grids will be considered in future work. Note that $\{k_j\}$ = set of all nodes connected to $j$ (but excluding $i$). Node 1 corresponds to the substation transformer at the point of common coupling (PCC). $\mathcal{N}$ and $\mathcal{E}$ are the sets of all nodes/buses and edges/branches in the network, respectively. The power flow model is specified by the following constraints:

\begin{align*}
    & v_j - v_i = (R_{ij}^2 + X_{ij}^2)l_{ij} -2(R_{ij}P_{ij} + X_{ij}Q_{ij}) \quad \forall (i,j) \in \mathcal{E}\\
    & P_{ij} = R_{ij}l_{ij} - P_j + \sum_{k\in \{k_j\}} P_{jk} \quad \forall (i,j) \in \mathcal{E}\\
    & Q_{ij} = X_{ij}l_{ij} - Q_j + \sum_{k\in \{k_j\}} Q_{jk} \quad \forall (i,j) \in \mathcal{E}\\
    & P_{ij}^2 + Q_{ij}^2 \leq v_il_{ij}, \quad P_{ij}^2 + Q_{ij}^2 \leq \overline{S}_{ij}^2, \quad 0 \leq l_{ij} \leq \overline{S}_{ij}^2/v_i \quad \forall (i,j) \in \mathcal{E}\\
    & P_j \in [\underline{P}_j,\overline{P}_j], \quad Q_j \in [\underline{Q}_j,\overline{Q}_j], \quad v_j \in [\underline{v}_j,\overline{v}_j] \quad \forall j \in \mathcal{N} \\
    & \text{where } l_{ij} = |I_{ij}|^2 \text{ and } v_i = |V_i|^2 \\
    & P_i(t) = P_i^{PV}(t) + P_i^{BS}(t) + P_i^{HP}(t) + P_i^{EV}(t) + P_i^L(t), \\
    & Q_i = Q_i^{PV} + Q_i^{BS} + Q_i^{EV}(t) + Q_i^L
\end{align*}

\noindent Note that throughout this paper, $P_i$ and $Q_i$ denote net active power injections (generation minus load) and net reactive power, respectively. Thus, $P_i, Q_i > 0$ implies net generation while net loads correspond to $P_i, Q_i < 0$. So, $P_i^{PV} \geq 0$ and $P_i^{HP}, P^L_i \leq 0$, whereas $P_i^{BS},P_i^{EV}$ are positive while discharging and negative while charging. We also add the following constraints:

\begin{align*}
    & \text{Transformer capacity constraints:} && P_{1j}^2 + Q_{1j}^2 \leq S_{tran}^2, \quad 0 \leq l_{1j} \leq \frac{S_{tran}^2}{v_{max}} \quad \forall j \in j_1 \\
    & \text{PCC power balance:} && P_{1} = \sum_{j \in {j_1}} P_{1j}, \quad Q_{1} = \sum_{j \in {j_1}} Q_{1j} \\
    & \text{Substation capacity (slack bus at node 1):} && \underline{P}^{PCC} \leq P_{1} \leq \overline{P}^{PCC}, \quad \underline{Q}^{PCC} \leq Q_{1} \leq \overline{Q}^{PCC}
\end{align*}

\noindent where $j_1$ denotes all the nodes connected directly to the PCC at node 1. These underlying constraints apply to both the static and dynamic cases. We can solve this feasibility problem (without any objective function) to check whether the DER power injections and dispatch results satisfy the power flow constraints.

\subsection{AC optimal power flow (ACOPF) for the dynamic case}

In the dynamic coordinated case, we solve the ACOPF problem, which minimizes the following objective function, subject to the power flow constraints.

\begin{align*}
    \min f^{OPF} & = \\
    & \text{minimize PV curtailment}   \\
    & \sum_{i \in bus} \sum_{t \in \mathcal{H}} \beta_{pv}  (P_i^{PV}(t) - \overline{P}^{PV}_i  \alpha^{PV}(t))^2 \\
    & \text{minimize thermal line losses}  \\
    & + \sum_{(i,j) \in \mathcal{E}} \sum_{t \in \mathcal{H}} l_{ij} R_{ij} \\
    & \text{maximize PV usage for BS and EV charging}   \\
    & + \sum_{i \in bus} \sum_{t \in \mathcal{H}} \left[\beta_{coupl,bs} \alpha^{PV}(t) P_i^{BS}(t) + \beta_{coupl,ev} \alpha^{PV}(t) P_i^{EV}(t)\right]  \\
    & \text{minimize absolute thermal discomfort}  \\
    & + \sum_{i \in bus} \sum_{t \in \mathcal{H}} \beta_T  (T_{i}^{in}(t) - T_{i}^*)^2 \\
    & \text{minimize HP and BS cycling}  \\
    & + \sum_{i \in bus} \sum_{t \in \mathcal{H}^-} \beta_{hp} (P_i^{HP}(t+1) - P_i^{HP}(t))^2 + \beta_{bs}  (P_i^{BS}(t+1) - P_i^{BS}(t))^2 \\
    & \text{minimize EV cycling and track desired SOC}  \\
    & + \sum_{i \in bus} \sum_{t \in \mathcal{H}^-} \beta_{ev1} (P_i^{EV}(t+1) - P_i^{EV}(t))^2 + \beta_{ev2}(SOC^{EV}_i(t^*) - SOC_i^{EV*})^2  \\
    & \text{minimize cost of power import at PCC}  \\
    &  + \sum_{t \in \mathcal{H}} \lambda(t) P_{1,t} - \sum_{i \in bus} \sum_{t \in \mathcal{H}} (\lambda(t) - \bar{\lambda})  P_i^{BS}(t)
\end{align*}

\noindent where $\mathcal{H}$ is the planning horizon (defined in \cref{sec:der_modeling}), $\mathcal{H}^-$ denotes $\mathcal{H}$ excluding its final time step (since the cycling terms involve forward differences), and $\alpha^{PV}(t)$ is the forecasted, time-varying solar irradiance profile. Note that the third term encourages battery charging during daylight hours when solar PV is available. The last term in the objective also encourages charging during low-price periods (when LMP $\lambda(t)$ is below average $\bar{\lambda}$), and discharging when LMP is high. In addition to ACOPF, the dynamic coordination problem also includes several more constraints (including intertemporal constraints) to represent each of the DER models. Note that adding such detailed DER models requires the use of integer variables, e.g., to prevent simultaneous BS charging and discharging, or to switch between HP heating and cooling modes. Thus, for the dynamic coordinated case, the overall optimization problem solved for each iteration (DER penetration level) is a mixed-integer second-order cone program (MISOCP).

Although $f^{OPF}$ has several terms, the problem is single-objective: the terms are combined in one weighted sum. The per-device objectives $f^{PV}_i$, $f^{BS}_i$, $f^{EV}_i$ and $f^{HP}_i$ written out next to the device models in \cref{sec:der_modeling} are not separate objectives; they are the components of $f^{OPF}$ associated with each device, and they use the same weights $\beta$. All penalty weights ($\beta_{pv}$, $\beta_T$, $\beta_{hp}$, $\beta_{bs}$, $\beta_{ev1}$, $\beta_{ev2}$) are set to 1 in this study, so the penalty terms are weighted equally (\cref{tab:hp-parameters}).

\subsection{DER modeling}
\label{sec:der_modeling}

Multiperiod optimization (MPO) accounts for all the device-level constraints of individual DERs, including time-coupled (or intertemporal) state constraints for the BS, EV, and HP. The simulation time step is $\Delta t = 5$ minutes, the total simulation time is $\mathcal{T} = [0,T]$ with $T = 24$ hours, and the planning horizon for the MPO is given by $\mathcal{H} = [t_H, t_H + (H-1)\Delta t] \subset \mathcal{T}$. We assume that the market-clearing time step is also equal to $\Delta t$, consistent with typical sub-hourly settlement intervals in US wholesale and distribution markets. Here, $H$ is the length (number of time steps) of the planning horizon, and we use a receding-horizon implementation in which the optimization is solved repeatedly as the simulation advances, with the horizon sliding forward at each step.

A 24-hour simulation horizon is appropriate for this planning-level HC study. How much hosting capacity storage enables depends on whether its energy capacity is sufficient to shift solar PV output and absorb demand peaks within a day; multi-day dynamics would be needed primarily for seasonal storage or hydrogen applications, which are outside our scope. The terminal SOC constraint ($SOC_i^{BS}(0) = SOC_i^{BS}(T)$) ensures that no day-to-day energy drift is introduced, making each simulated day representative of steady-state operation. Extending to multi-day horizons would provide richer seasonal variation but at substantially higher computational cost; we consider this a natural direction for future work.

\subsubsection{PV model}

The maximum PV generation output is determined by the forecasted, time-varying solar irradiance profile $\alpha^{PV}(t)$ along with its maximum rated capacity $\overline{P}^{PV}_i$, which can be curtailed if needed.
\begin{gather*}
    0 \leq P^{PV}_i(t) \leq \alpha^{PV}(t) \overline{P}^{PV}_i
\end{gather*}
The corresponding term of $f^{OPF}$ penalizes curtailment of available PV output:
\begin{gather*}
    f^{PV}_i = \beta_{pv} \sum_{t=t_H}^{t_H + (H-1)\Delta t} (\alpha^{PV}(t) \overline{P}^{PV}_i - P^{PV}_i(t))^2
\end{gather*}
This term is not a separate PV objective. It is the PV component of $f^{OPF}$, and it is zero whenever PV is not curtailed, as in all hosting capacity results reported here.

\noindent While the active power output of a PV unit is non-dispatchable, its reactive power injection is determined by variable power factor control enabled by smart inverters. We only consider conventional grid-following inverters for PV, BS, and EVs since these are most common today. Grid-forming inverters can also be considered in future work \cite{dorfler2023control}.
$$
-P_i^{PV} \tan(\cos^{-1}(\underline{pf})) \leq Q_i^{PV} \leq P_i^{PV} \tan(\cos^{-1}(\underline{pf}))
$$

\subsubsection{BS model}

$$P^{BS}_i = P^{BS, d}_i - P^{BS, c}_i \quad P^{BS, d}_i, P^{BS, c}_i \geq 0$$ where $P^{BS, d}_i$ and $P^{BS, c}_i$ denote the discharging and charging power injections, respectively. We introduce binary variables $u^{BS,d}_i, u^{BS,c}_i \in \{0,1\}$ to denote whether the BS is discharging ($u^{BS,d}_i = 1, u^{BS,c}_i = 0$) or charging ($u^{BS,d}_i = 0, u^{BS,c}_i = 1$). Simultaneous charge and discharge are prevented by the following constraint: $u^{BS,d}_i + u^{BS,c}_i \leq 1$. The state of charge (SOC) dynamics of the battery are:
\begin{gather*}
    SOC^{BS}_i(t+1) = (1-\delta_{BS}^i)SOC^{BS}_i(t) + \frac{\Delta t}{\overline{E}^{BS}_i} \left(P^{BS^c}_i(t) \eta^{BS}_i - \frac{P^{BS^d}_i(t)}{\eta^{BS}_i}\right)\\
    0 \leq P^{BS,d}_i(t) \leq u^{BS,d}_i\overline{P}^{BS}_i, \quad 0 \leq P^{BS,c}_i(t) \leq u^{BS,c}_i\overline{P}^{BS}_i \\
    \underline{SOC}_i^{BS} \leq SOC^{BS}_i(t) \leq \overline{SOC}^{BS}_i, SOC_i^{BS}(0) = SOC_i^{BS}(T)
\end{gather*}

\noindent where $\delta^{BS}_i, \eta^{BS}_i$ and $\overline{E}^{BS}_i$ are the BS self-discharge rate, charge-discharge efficiency, and maximum energy capacity, respectively. We also have a terminal constraint to ensure that the state of charge at the start and end of the simulation must be equal. Note that generally, $\underline{P}_i^{BS} = -\overline{P}^{BS}_i$ since we assume the BS has the same charging and discharging power capabilities.
During BS operation, we aim to minimize the cycling cost to avoid excessive charge and discharge cycles, which can shorten the battery's lifetime. This cycling penalty serves as a proxy for battery degradation: by penalizing rapid power fluctuations, it discourages high-rate cycling that accelerates capacity fade in practice. A more detailed degradation model (e.g., cycle counting with a rainflow algorithm or an electrochemical model) could be incorporated in future work, but the cycling cost captures the dominant driver of degradation at the planning level.
\begin{gather*}
    f^{BS}_i(P^{BS}_i) = \beta_{bs} \sum_{t=t_H}^{t_H + (H-1)\Delta t} (P^{BS}_i(t+1) - P_i^{BS}(t))^2
\end{gather*}
where we sum over all the time steps in the planning horizon starting at time $t_H$. We also have variable power factor control for battery reactive power:
\begin{align*}
    & -P_i^{BS, d \; | \;c} \tan(\cos^{-1}(\underline{pf})) \leq Q_i^{BS, d \; | \;c} \leq P_i^{BS, d \; | \;c} \tan(\cos^{-1}(\underline{pf}))
\end{align*}

\subsubsection{EV model}

The EV battery has dynamics and constraints similar to those described above. In addition, we place restrictions on EV availability. We assume the EV is not present at the building or home during the period $[t_1,t_2]$, e.g., between 9am and 5pm when the owner is at work. This is enforced by $P_i^{EV}(t) = 0 \; \forall \; t \in [t_1, t_2]$. These times may vary across different nodes. We can impose a similar cycling cost on the EV to extend its lifetime. We also add a tracking objective to capture the EV owner's preference for achieving a certain desired SOC ($SOC_i^*$) by a specific time ($t^*$); for example, the owner may need the EV to be 90\% charged by 9am before work. Thus, the EV objective function is
\begin{gather*}
    f^{EV}_i = \beta_{ev1} \sum_{t=t_H}^{t_H + (H-1)\Delta t} (P^{EV}_i(t+1) - P_i^{EV}(t))^2 + \; \beta_{ev2} (SOC^{EV}_i(t^*) - SOC^{EV^*}_i)^2
\end{gather*}
We assume that all EVs are capable of vehicle-to-grid (V2G), allowing bidirectional power flow. This means that in addition to managed charging, they can also discharge and inject power back into the grid while connected to their charger. For simulations, we assume that the homes are equipped with a 9.6 kW EV charger, which is the most common level 2 residential charging setup in the US. We also assume a 67 kWh energy storage capacity for EVs, allowing for full EV charging in about 6-8 hours, compared to a 3-4 h duration for residential batteries. We model only EVs that belong to homes on the feeder. Charging by EVs that arrive from outside the feeder, for example at workplace or public chargers, is not represented as a flexible resource here; it would add daytime load and could also offer additional midday flexibility, and we leave it to future work.

\subsubsection{HVAC or heat pump model}

The HVAC dynamics describe how the power drawn affects the indoor air temperature in the home or building. In this work, we consider the HVAC unit to be a heat pump, which can serve as either a heating or cooling device depending on the ambient temperature. These thermostatically controlled loads (including heat pumps, HVAC units, and water heaters) can provide load flexibility by modifying their temperature setpoints \cite{Sanandaji2016RampingLoads}. If $T^{out}_i(t) > T^{in}_i(t)$, the temperature dynamics of the HP in cooling mode (i.e., when it acts as an air conditioner) are (\cite{Zhao2017ALoads}):
\begin{gather*}
    T^{in}_i(t+1) = \theta_i T_i^{in}(t) + (1-\theta_i)\left(T^{out}_i(t) + \rho_i P^{HP}_i(t)\right)
\end{gather*}
where $\theta_i = e^{\frac{-\Delta t}{R^{th}_iC^{th}_i}} \approx 1 - \frac{\Delta t}{R^{th}_iC^{th}_i}, \rho_i = R^{th}_i \eta_i$ and $R^{th}_i, C^{th}_i, \eta_i$ are the equivalent thermal resistance, thermal capacitance, and coefficient of performance of the system, respectively. The thermal parameters for a residential (single-family) HP unit are given by \cref{tab:hp-parameters} \cite{Hao2015AggregateLoads}. The temperature dynamics in heating mode, when $T^{out}_i(t) < T^{in}_i(t)$, are:
\begin{gather*}
    T^{in}_i(t+1) = \theta_i T_i^{in}(t) + (1-\theta_i)\left(T^{out}_i(t) - \rho_i P^{HP}_i(t)\right)
\end{gather*}
The HP operation is also subject to operational limits on power draw and indoor temperature:
\begin{gather*}
    \underline{P}^{HP}_i \leq P^{HP}_i(t) \leq 0, \underline{T}_i^{in} \leq T^{in}_i(t) \leq \overline{T}^{in}_i
\end{gather*}
Note that here the lower limit is determined by the maximum rated power consumption of the HVAC unit, i.e., $\underline{P}^{HP}_i = -P_r^{HP}$ since the HP always acts as a load. For the HP objective, we would like to track a desired temperature setpoint to maximize the thermal comfort of occupants and avoid excessive cycling between heating and cooling modes (or ON-OFF):
\begin{gather*}
    f^{HP}_i = \beta_T \sum_{t=t_H}^{t_H + (H-1)\Delta t} (T^{in}_i(t) - T^{in^*}_i)^2 + \beta_{hp} \sum_{t=t_H}^{t_H+(H-1)\Delta t} (P_i^{HP}(t+1) - P_i^{HP}(t))^2
\end{gather*}
In addition to HP units, using a similar approach, we may also consider other types of thermostatically controlled loads (TCLs) such as water heaters (WHs). We assume HPs do not inject or absorb any reactive power, i.e., $Q_i^{HP}=0$. In order to make the model more realistic, we place explicit upper limits on the ramp rate or change in HP consumption between time steps, where $\bar{\Delta}_{hp}$ is the upper bound as a fraction of the rated HP power, e.g., 30\% (assumed in our study).
$$
\|P_i^{HP}(t+1) - P_i^{HP}(t)\|_2 \leq \bar{\Delta}_{hp} P_r^{HP}
$$

\begin{table}
    \centering
    \begin{tabular}{llll}
      \toprule
      Symbol & Description & Value & Unit \\
      \midrule
      \multicolumn{4}{l}{\textit{Heat pump (per home) \cite{Hao2015AggregateLoads}}} \\
      $C^{th}_i$ & thermal capacitance & 2 & $\mathrm{kWh} /{ }^{\circ} \mathrm{C}$ \\
      $R^{th}_i$ & thermal resistance & 2 & ${ }^{\circ} \mathrm{C} / \mathrm{kW}$ \\
      $P_r^{HP}$ & rated electrical power$^{a}$ & 3--5 & kW \\
      $\eta_i$ & coefficient of performance & 2.5 & -- \\
      $T^{in^*}_i$ & indoor temperature setpoint & 22.5 & ${ }^{\circ} \mathrm{C}$ \\
      $\bar{\Delta}_{hp}$ & ramp limit (fraction of $P_r^{HP}$) & 0.3 & -- \\
      \multicolumn{4}{l}{\textit{EV (per home) and storage}} \\
      $P_r^{EV}$ & EV charger rating & 9.6 & kW \\
      $E_r^{EV}$ & EV battery capacity & 67 & kWh \\
      $E_r^{BS}$ & storage duration (energy/power) & 3 & h \\
      \multicolumn{4}{l}{\textit{Optimization}} \\
      $\beta_{pv}, \beta_T, \beta_{hp}, \beta_{bs}, \beta_{ev1}, \beta_{ev2}$ & penalty weights in $f^{OPF}$ & 1 & -- \\
      $\Delta t$ & time step & 5 & min \\
      \bottomrule
    \end{tabular}
    \caption{Model parameters. The outdoor temperature is a scenario input (\cref{tab:input_data}), not a fixed parameter. $^{a}$3~kW per unit in the deterministic heat pump case study (\cref{sec:res_hp}) and 5~kW in the other runs; the thermal parameters are those of a 5.6~kW reference unit \cite{Hao2015AggregateLoads}.}
    \label{tab:hp-parameters}
\end{table}

\subsection{Deterministic iterative (DI) approach for HCA}

\begin{figure}
    \centering
    \includegraphics[width=\textwidth]{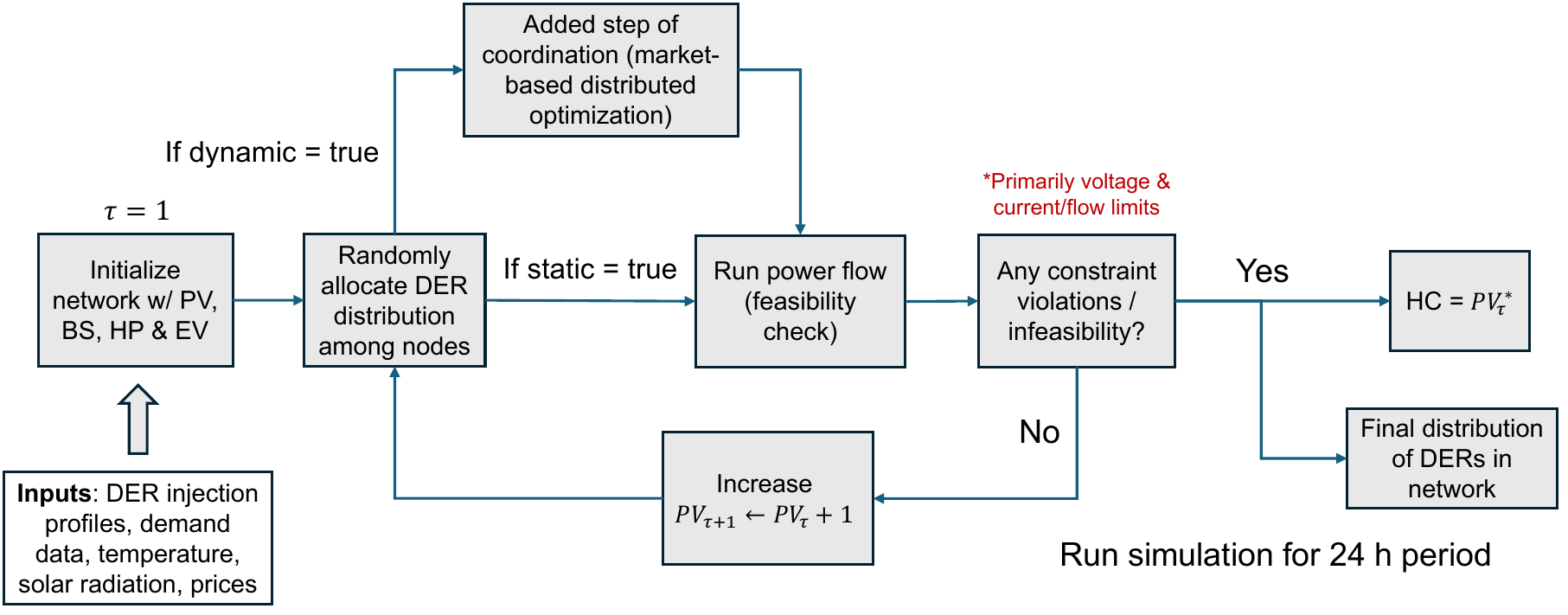}
    \caption{Deterministic iterative HCA: DER penetration is raised in small increments and the static and dynamic models are solved at each step until power flow becomes infeasible; the last feasible step is the hosting capacity.}
    \label{fig:det_iter}
\end{figure}

We first consider a deterministic case where we do not account for uncertainty and only consider a single scenario of inputs. One simplification is to only restrict ourselves to a specific time of day, which corresponds to the worst possible inputs. For example, if we are mainly focused on PV hosting capacity, we can consider just mid-day hours (e.g., 12-1pm) during peak PV output since the main concern is overvoltage due to excess generation \cite{Torquato2018ASystems}. However, while estimating HC for other DERs like EVs, BS, or HPs, we may also need to include other time periods since there are also risks of undervoltage (due to excess load) or line overloading. As summarized in \cref{fig:det_iter}, we implement an iterative approach where we gradually increase the DER penetration in small increments and run both the static and dynamic models until we hit power flow infeasibility. This allows us to keep track of maximum bus voltages and line loadings at each intermediate penetration level. Note that most papers based on iterative approaches focus on estimating the HC of only one type of DER at a time, e.g., maximizing PV only or HP only \cite{Zhan2022TowardsCapacity}. However, the iterative approach is relatively more computationally expensive than the optimization-based methods described below, and may not give as granular HC estimates due to the fixed step size increments at each iteration.

\subsection{Non-deterministic case}

In the static case, we have stochasticity in the baseline power profiles of DERs (BS/EV/HP) that we assume for the simulation, along with inherent uncertainty in solar output, fixed load consumption, and prices. In the dynamic case, we have uncertainty in temperature (affecting HP load and flexibility), fixed load profiles, solar radiation, and prices. In addition, there is stochasticity in the time availability windows of EVs \cite{CastilloJr.2022UsingCommercialization}. It is reasonable to assume that all these variables can be treated as roughly independent random variables. Thus, we can randomly sample and generate each variable separately and then combine them to create each scenario. This input uncertainty captures both forecast (epistemic) uncertainty and the inherent variability or aleatoric uncertainty arising from factors like seasonal weather patterns \cite{Mulenga2020AGrids}. Grid planners need to accurately account for uncertainty for more realistic, conservative HC estimates, which are generally lower than deterministic results.

\begin{figure}
    \centering
    \includegraphics[width=\textwidth]{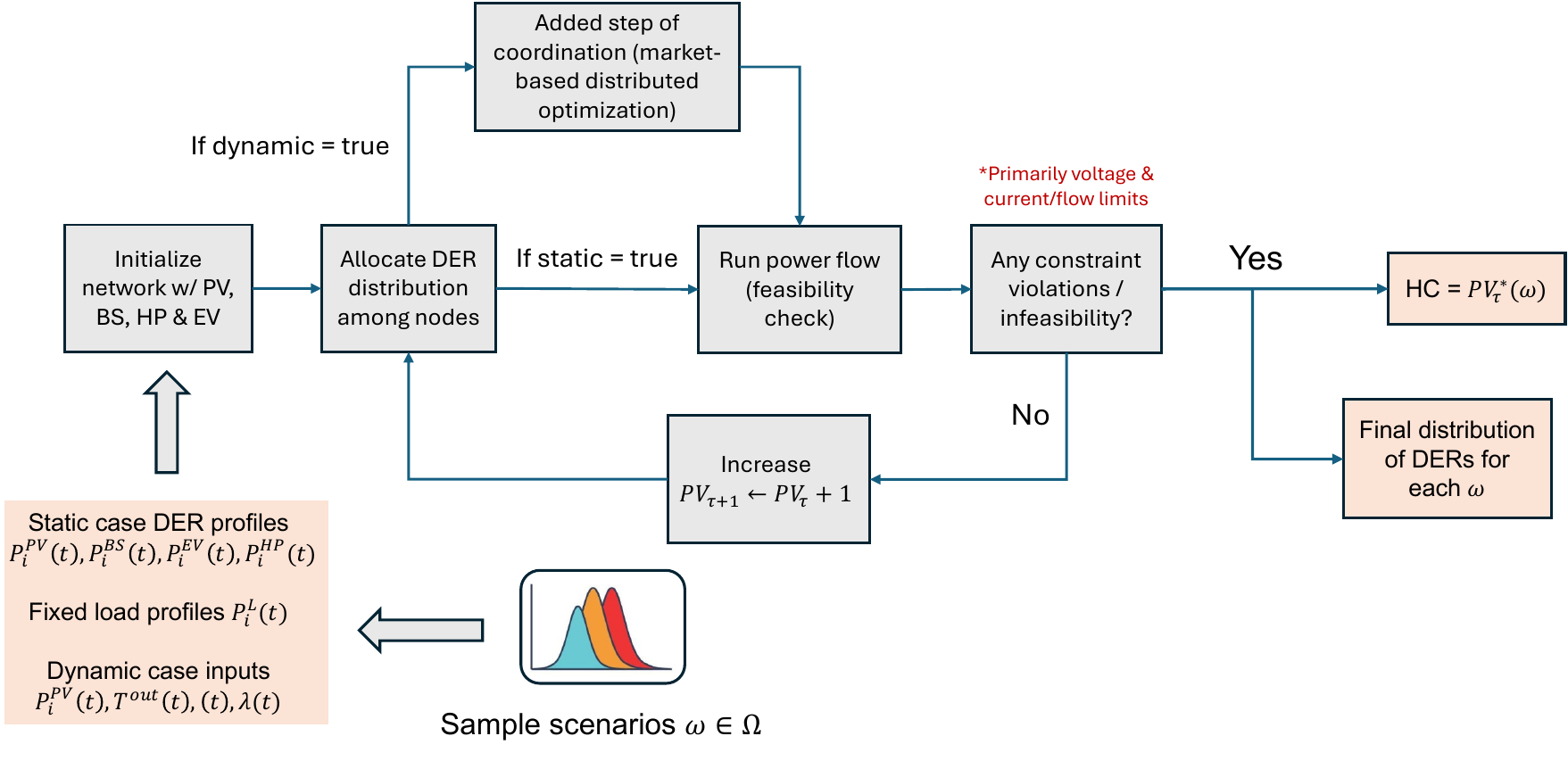}
    \caption{Stochastic iterative HCA: the deterministic iterative method is repeated for many sampled input scenarios to obtain the distribution of hosting capacity and grid metrics.}
    \label{fig:stoch_iter}
\end{figure}

This is done by repeatedly running the iterative method for many scenarios of time series profiles, in order to obtain distributions of results across scenarios. Thus, we convert the deterministic iterative method to a stochastic iterative method using Monte Carlo sampling. The approach is summarized in \cref{fig:stoch_iter} and consists of the following steps.

\begin{enumerate}
    \item We generate many scenarios or realizations of all input time series profiles by adding Gaussian noise to baseline profiles. These profiles were derived from real-world historical data from various sources, such as the Pecan Street smart meter dataset \cite{pecanstreet_dataport}.
    \item We then run both the static and dynamic deterministic iterative HCA by considering one scenario at a time. Since the scenarios can all be run independently, we can parallelize these to significantly speed up the simulation.
    \item Pick the lowest HC over all the scenarios (in order to be conservative) and also record other metrics like the mean and median. We can then plot the estimated HCs as distributions and obtain their kernel density estimates as well. Quantifying uncertainty in this manner also provides valuable information for grid planners, operators, and load-serving entities.
\end{enumerate}

Several frameworks can handle uncertainty in HCA. Stochastic programming (SP) optimizes the expected outcome over a finite set of sampled scenarios and requires feasibility in each of them \cite{ConejoDecisionMarkets}. Robust optimization (RO) requires feasibility for every realization in an uncertainty set and is typically more conservative \cite{bertsimas2004price}; information gap decision theory (IGDT) protects against the worst realization when distributions are unknown \cite{vahid2023hybrid,liao2022information}; and chance constraints bound the probability of violating each limit but need accurate distributions and can be hard to compute. Because our inputs are derived from historical profiles from which scenarios can be sampled directly, we use stochastic programming, which balances optimality, feasibility and tractability for this application.

\subsection{Two-stage stochastic program (2-SSP) for HCA}

Traditional HC analysis determines a scalar quantity: the maximum amount of a given DER type that can be added at a location before a constraint is violated. The 2-SSP extends this concept to a \textit{joint HC planning problem} that simultaneously optimizes the siting (which nodes) and sizing (how much capacity) of multiple DER types, while ensuring power flow feasibility across a set of representative operating scenarios. Rather than a single HC value for a specific DER, the output is a feasible penetration envelope --- the set of jointly achievable DER penetration combinations --- which captures the trade-offs and complementarities among different resource types. This is a richer and more actionable planning tool than a scalar HC estimate, and it directly links the design-stage investment decision (1st stage) with operational feasibility (2nd stage).

Planning and operation are combined because, with coordination, the extra hosting capacity is created by the dispatch policy. A PV system that would violate voltage limits under static operation becomes feasible because storage and heat pumps nearby are dispatched to absorb its output. A siting decision made without seeing that dispatch would either be infeasible in operation or leave capacity unused. The 2-SSP therefore lets the first-stage planning decision anticipate the second-stage coordinated operation in every scenario, which is what makes the co-optimized siting in \cref{sec:results_2ssp} possible.

A two-stage stochastic program (2-SSP) has a first (planning) stage, whose decisions are made before the uncertainty is realized, and a second (operational or recourse) stage with one subproblem for each scenario or realization. We combine both the 1st and 2nd stages into a single large-scale deterministic equivalent program. We then solve for separate sets of 2nd-stage decision variables for each scenario, and the goal is to obtain solutions for the 1st stage decisions that are also feasible for the 2nd stage. This requires coupling between the two stages. Although such problems are infinite dimensional due to continuous probabilities, we use a discretized version of the 2-SSP to get a finite optimization problem, also known as sample average approximation (SAA). Sampling a large number of diverse scenarios (with Monte Carlo simulations) helps ensure that the finite case is still a good approximation for the infinite case.

The HCA can be written as a 2-stage stochastic optimization problem with fixed recourse. In the 1st stage, we choose the overall DER penetrations before scenario realization, and in the 2nd stage (one recourse problem per scenario), we run power flow simulations for all realized scenarios. The 1st stage decisions are the DER penetrations along with their locations and capacities at nodes. The 2nd stage problem is an ACOPF model that checks feasibility for each scenario $\omega \in \Omega$. Thus, the 1st stage is concerned with design, planning, and investment --- the grid planner has to make these decisions before seeing the input uncertainty. The 2nd stage is concerned with actual grid operation to set the power flow solutions and DER dispatch. These can only be decided after the uncertain input scenarios $\omega$ are realized. This results in a mixed-integer second-order cone program whose continuous relaxation is convex. The advantages of such an approach include (i) better computational efficiency, especially when considering non-deterministic problems over longer time horizons, and (ii) more flexibility for the decision-maker.

\subsubsection{First stage: Planning and design problem}

Here we aim to maximize the capacities of PV, EV, and HP, for a given BS level ($\overline{P}^{BS} = \sum_i \overline{P}^{BS}_i$). We consider the \textit{total} maximum BS capacity (both rated power and energy storage capacity) as exogenous variables or inputs (specified by the planner, utility, or grid operator). This is because we can generally arbitrarily increase BS levels by always controlling whether to charge, discharge, or turn off batteries as needed, depending on grid conditions. Thus, we do not define the notion of HC for batteries. Storage therefore acts throughout this paper as an enabler of PV and heat pump hosting capacity: storage penetrations reported in the results are inputs chosen by the planner, not hosting capacities. Higher BS penetrations would allow us to increase dynamic HC further, but come at an added cost. While the total BS penetration across the entire feeder is given, its distribution among the nodes is determined optimally using our program.

We treat EVs and HPs as DERs rather than passive loads because their consumption is flexible and dispatchable: both can shift their energy demand in time (EVs by scheduling charging within availability windows; HPs by adjusting temperature setpoints within comfort bounds), and EVs additionally support bidirectional power flow via V2G. This flexibility makes them active participants in grid optimization, not merely additional load, and it is this dispatchable character that distinguishes them from conventional fixed loads in our framework.

The decision variables represent the penetration levels of each DER type at each node. In the case of PV and BS, $x_i^{PV}$ and $x_i^{BS}$ refer to the penetration of PV capacity and BS relative to the total baseline or nominal load at that node $i$ (as specified by the IEEE datasheet). In the case of EVs and HPs, the decision variables $x_i^{EV}, x_i^{HP}$ represent the proportion (\%) of homes or buildings (connected to node $i$) that have been electrified with EVs and HPs, respectively. The objective function is to maximize the total DER capacity. The first stage problem is formulated as:

\begin{gather}
    \max_{x_i} f^I = \sum_i \left( x_i^{PV} + x_i^{BS} + x_i^{EV} + x_i^{HP} \right) \label{eq:stage1_revised}\\
    \text{s.t. } 0 \leq x_i^{PV} , x_i^{BS} \quad  0 \leq x_i^{EV}, x_i^{HP} \leq 1, \quad \sum_i x_i^{BS} \overline{P^L}_{nom,i} \leq \overline{x}^{BS} \overline{P^L}_{nom} \label{eq:stage1_storage_constr}
\end{gather}

\noindent where the constraints are upper and lower limits on the penetrations of each DER type at each node and the last constraint in \cref{eq:stage1_storage_constr} restricts the maximum BS capacity to the level pre-determined by the planner, as a fraction of the total baseline feeder load $\overline{P^L}_{nom}=\sum_i \overline{P^L}_{nom,i}$. Note that $x_i^{PV}, x_i^{BS}$ are not upper bounded but $x_i^{HP}, x_i^{EV}$ are at most 1, which would mean that 100\% of homes have been electrified. The objective is a single, unweighted sum of nodal penetrations, so the first stage is not a multi-objective problem. Because the total storage is fixed by \cref{eq:stage1_storage_constr}, the $x_i^{BS}$ term only ensures that the storage budget is used; its allocation across nodes is chosen to make the other penetrations feasible. The optimal combination of DER types is generally not unique, since many siting patterns can achieve the same total. We therefore study trade-offs among DER types by fixing the storage and heat pump levels and re-solving for the maximum PV penetration (\cref{fig:2ssp_sensitivity}), rather than by computing a Pareto front. Solving the optimization problem in \cref{eq:stage1_revised} also solves for the optimal siting and sizing of DERs. Instead of randomly allocating DERs over the network, given any desired (feasible) DER penetration level, this will optimally determine the nodal locations of different types of DERs, along with their associated capacity.

\subsubsection{Coupling between the 1st and 2nd stages}

In this study, we assume that the network topology as well as the distribution of houses or buildings over the feeder are fixed. The number of houses at each node $i$ is given by $n_i^h$. For simplicity, we also assume the same standardized sizes of EV and HP units for each home. $P_r^{EV}$ and $E_r^{EV}$ denote the common rated power and energy storage capacities of EVs, and $P_r^{HP}$ is the standard HP rating. The total capacities of the DERs at each node are:

\begin{gather*}
    \overline{P}_i^{HP} = x_i^{HP} n_i^h P_r^{HP}, \quad \overline{P}_i^{EV} = x_i^{EV} n_i^h P_r^{EV}, \quad \overline{E}^{EV}_i = x_i^{EV} n_i^h E_r^{EV} \\
    \overline{P}_i^{BS} = x_i^{BS} \overline{P^L}_{nom,i} \quad \overline{E}^{BS}_i = \overline{P}_i^{BS} E_r^{BS}, \quad \overline{P}^{PV}_i = x_i^{PV} \overline{P^L}_{nom,i}
\end{gather*}

\noindent Thus, the 1st stage decision variables $\textbf{x}$ couple the two stages since they set the maximum DER capacities at each node, which in turn constrain the 2nd stage problem.

\subsubsection{Second stage problem: Operation}

The 2nd-stage formulation differs between the static and dynamic cases.

\textit{Static case.} In the static case, we assume that all the DER injections are inflexible. Thus, we only solve a feasibility problem to check if the given DER injections satisfy all the power flow constraints. In this case, the scenarios consist of different time series profiles for inflexible DER injections, including HP load consumption, EV/BS charge-discharge cycles, fixed loads, and PV radiation. There are a total of 5 random variables, and we assume these to be independent, but not necessarily identically distributed. We directly generate multiple scenarios by randomly perturbing or adding noise to baseline or nominal profiles derived from actual historical data.

Thus, the 2-stage static SP is a recourse problem with only a feasibility check in the 2nd stage. The feasibility of first-stage decisions under uncertainty is assessed in the second stage. The 2nd stage has a zero objective function, $f^{II} = 0$, and the decision variables are $\mathbf{y} = \{P_i(t), Q_i(t), P_{ij}(t), Q_{ij}(t),l_{ij}(t),v_i(t)\} \; \forall t \in \mathcal{T}$, with a separate set of variables for each scenario $\omega_k \in \Omega$. Note that the 2nd stage is a multiperiod optimization problem. The overall static 2-SSP problem is then given by:

\begin{align}
    \max_{\mathbf{x},\mathbf{y}_1, \dots \mathbf{y}_K} \quad & f^I(\mathbf{x}) \\
    & \text{s.t.: } \text{Constraints on } \mathbf{x} \text{ in \cref{eq:stage1_revised}}  \nonumber\\
    & h_{pf}(\omega_k,\mathbf{x},\mathbf{y}_k, t) = 0, \quad g_{pf}(\omega_k,\mathbf{x},\mathbf{y}_k, t) \leq 0, \quad k = 1, 2, \dots K  \nonumber\\
    & P_i(\omega_k,t) = P^{PV}_i(\omega_k,t) + P^{EV}_i(\omega_k,t) + P^{BS}_i(\omega_k,t) + P^{HP}_i(\omega_k,t) + P^L_i(\omega_k,t)  \nonumber\\
    & Q_i(\omega_k,t) = Q^L_i(\omega_k,t) \; \forall t \in \mathcal{T}  \nonumber
\end{align}

\noindent where $h_{pf},\; g_{pf}$ are the AC power flow equality and inequality constraints, respectively, and $K$ is the total number of sampled scenarios. $\omega_k$ denotes the data for each scenario. In case we allow solar PV curtailment, we would have an additional decision variable for the PV power injection $P_i^{PV}$. Note that here $P^{EV}_i(\omega_k,t), P^{HP}_i(\omega_k,t), \text{ and } P^{BS}_i(\omega_k,t)$ are pre-specified inflexible stochastic profiles, and none of the DERs can provide reactive power support.

\textit{Dynamic case.} During the second stage with dynamic coordination, we solve the multiperiod optimal power flow problem that determines the dispatch after the realization of uncertain scenarios. This involves solving the ACOPF problem with the specified objective and all constraints corresponding to power flow and individual DER models, to find feasible solutions for each scenario $\omega \in \Omega$. Each scenario introduces uncertainty in the following inputs:

\begin{gather*}
    P_i^{PV}(\omega) = \overline{P}^{PV}_i \alpha(\omega) \quad \text{(without curtailment)} \\
    P_i^{PV}(\omega) \leq \overline{P}^{PV}_i \alpha(\omega) \quad \text{(with curtailment)} \\
    T_i^{out} = T_i^{out}(\omega) \quad \text{(assume common outdoor temp for all nodes)} \\
    \lambda = \lambda(\omega), \quad P_i^L = P_i^L(\omega)
\end{gather*}

\noindent Here $\alpha(\omega)$ is the stochastic PV output based on solar irradiation; $T(\omega)$ and $\lambda(\omega)$ are the external temperature profile and LMP, respectively. These three variables depend on the scenario but are the same for all nodes. The stochastic fixed load profiles $P_i^L(\omega)$ differ for each node and scenario. The 2nd stage decision variables are $\mathbf{y} = \{P_i(t), Q_i(t), P_{ij}(t), Q_{ij}(t),l_{ij}(t),v_i(t)\}$ along with controllable states $\{T_i^{in}(t), SOC^{BS}_i(t), SOC^{EV}_i(t)\}$ and DER-specific injections given by $\{P^{PV}_i, P_i^{BS}(t), P^{EV}_i(t), P^{HP}_i(t), Q^{PV}_i(t), Q^{BS}_i(t), Q^{EV}_i(t)\}$. The 2nd stage objective function is the ACOPF cost function $f^{II} = f^{OPF}$. To illustrate its structure, we first write this 2-stage SP in nested form, with a first-stage problem and a second-stage value function $V(\mathbf{x},\omega_k)$ for each scenario. The separable objective function is split into (i) a deterministic term representing decisions at the design stage and (ii) the expectation of a stochastic term which depends on the realization of uncertain parameters at the operation stage \cite{Zhou2013ASystems}. We use a discrete approach with sample average approximation (SAA), by replacing the expectation with a sample average over the scenarios. The first-stage problem of the dynamic 2-SSP is then given by:
\begin{align}
    \min_{\mathbf{x}} \quad & -f^I(\mathbf{x}) + \sum_k p_k V(\mathbf{x},\omega_k) \\
    & \text{s.t.: } \text{Constraints on } \mathbf{x} \text{ in \cref{eq:stage1_revised}} \nonumber
\end{align}

\noindent where $p_k$ is the probability of scenario $\omega_k$. The first-stage objective $f^I$ in \cref{eq:stage1_revised} is a total penetration to be maximized, whereas $f^{II}$ is an operating cost to be minimized, so $f^I$ enters the minimization with a negative sign. We assume that all scenarios are equally likely, i.e., $p_k = \frac{1}{K}$. However, it is also straightforward to use scenario-specific probabilities if we have access to the accurate probability distribution of the inputs over possible scenarios. The second-stage problem for scenario $\omega_k$ is:
\begin{align}
    V(\mathbf{x},\omega_k) & = \min_{\mathbf{y}_k} \quad f^{II}(\mathbf{x},\mathbf{y},\lambda_k) \\
    & \text{s.t.: } h_{pf}(\omega_k,\mathbf{x},\mathbf{y}_k,t) = 0, \quad g_{pf}(\omega_k,\mathbf{x},\mathbf{y}_k,t) \leq 0, \quad k = 1, 2, \dots K \nonumber\\
    & P_i(\omega_k,t) = P^{PV}_i(\omega_k,t) + P^{EV}_i(\omega_k,t) + P^{BS}_i(\omega_k,t) + P^{HP}_i(\omega_k,t) + P^L_i(\omega_k,t) \nonumber\\
    & Q_i(\omega_k,t) = Q^{PV}_i(\omega_k,t) + Q^{EV}_i(\omega_k,t) + Q^{BS}_i(\omega_k,t) + Q^L_i(\omega_k,t) \nonumber\\
    & h_{der}(\omega_k,\mathbf{x},\mathbf{y}_k,t) = 0, \quad g_{der}(\omega_k,\mathbf{x},\mathbf{y}_k,t) \leq 0, \quad k = 1, 2, \dots K \nonumber
\end{align}

\noindent Note that the second-stage problem is multiperiod and at each time step, it is repeatedly solved for a finite time horizon into the future $\mathcal{H}$, in order to cover the whole simulation period. By combining the two stages into a single large-scale optimization problem, we obtain a single-level large-scale \textit{deterministic equivalent} of this nested two-stage program. The overall optimization problem is given by:

\begin{align}
    \min_{\mathbf{x},\mathbf{y}_1, \dots \mathbf{y}_K} \quad & -f^I(\mathbf{x}) + \sum_{k=1}^K p_k f^{II}(\mathbf{x},\mathbf{y}_k,\lambda) \\
    & \text{s.t.: } \text{Constraints on } \mathbf{x} \text{ in \cref{eq:stage1_revised}} \nonumber\\
    & h_{pf}(\omega_k,\mathbf{x},\mathbf{y}_k,t) = 0, \quad g_{pf}(\omega_k,\mathbf{x},\mathbf{y}_k,t) \leq 0, \quad k = 1, 2, \dots K \nonumber \\
    & P_i(\omega_k,t) = P^{PV}_i(\omega_k,t) + P^{EV}_i(\omega_k,t) + P^{BS}_i(\omega_k,t) + P^{HP}_i(\omega_k,t) + P^L_i(\omega_k,t) \nonumber\\
    & Q_i(\omega_k,t) = Q^{PV}_i(\omega_k,t) + Q^{EV}_i(\omega_k,t) + Q^{BS}_i(\omega_k,t) + Q^L_i(\omega_k,t) \nonumber\\
    & h_{der}(\omega_k,\mathbf{x},\mathbf{y}_k,t) = 0, \quad g_{der}(\omega_k,\mathbf{x},\mathbf{y}_k,t) \leq 0, \quad k = 1, 2, \dots K \nonumber
\end{align}
where $h_{der},\; g_{der}$ are the equality and inequality constraints for the DER models and state dynamics, respectively. The second-stage variables $\mathbf{y}_k$ form a separate block for each scenario $\omega_k$, but all blocks are optimized jointly with $\mathbf{x}$ in this single problem, so no iteration between a master problem and subproblems is needed. Note that in the above formulation, we used the mean (or expectation) of the 2nd stage cost as our risk measure. In order to be more conservative and risk-averse, we could instead use other convex risk measures like the conditional value at risk (CVaR) or expected shortfall \cite{Tan2017ApplicationResponse}. CVaR quantifies the expected loss in the worst-case scenarios beyond a specified threshold, thus better capturing tail risk.

\subsubsection{Enforcing DER model constraints}

An important consideration for the stochastic dynamic case is that we enforce DER model state constraints only at nodes where they are actually present. Such conditional constraints are non-convex but can be implemented as a mixed-integer program using big-M constraints, and then solved using solvers like Gurobi. Further details can be found in Appendix~\ref{app:der_presence}.

\subsection{Solving 2-stage stochastic problems}

Large two-stage stochastic programs are usually solved by decomposition: Benders decomposition and its L-shaped variant \cite{Benders1962,VanSlyke1969}, their integer and generalized extensions \cite{Laporte1993,Geoffrion1972}, scenario decomposition \cite{Rockafellar1995}, or, for multistage problems, stochastic dual dynamic programming (SDDP) \cite{Zou2019StochasticProgramming}. These cutting-plane methods rely on a convex second-stage value function. In our problem the second stage contains binary variables (battery charge/discharge and heat pump mode), so its value function is nonconvex and the standard cuts are not valid.

We therefore solve the deterministic equivalent directly, implemented in Julia using the JuMP package. Gurobi is able to efficiently solve such MISOCPs to guarantee global optimality as long as the problem is feasible, even though the optimal solution may not be unique (due to the presence of binary variables). While this approach has worked reasonably well for our moderately sized problems, it may not be computationally practical for larger-scale problems, with many more decision variables and constraints. In future work while considering larger networks with more scenarios, we will explore specialized solution algorithms that can exploit the two-stage, scenario-separable and temporal structure of the problem.

\subsection{Accelerating 2-SSP for HCA}
\label{sec:accel_2ssp}

One approach to improve the computational efficiency of this manual sweep approach is to reduce the number of scenarios that need to be run. We can use unsupervised machine learning methods to identify the most representative/characteristic scenarios and only run the simulation for these. This reduced scenario set must still include the most severe or extreme cases that will place more stress on the grid and make it challenging to satisfy constraints.

\begin{figure}
    \centering
    \includegraphics[width=\textwidth]{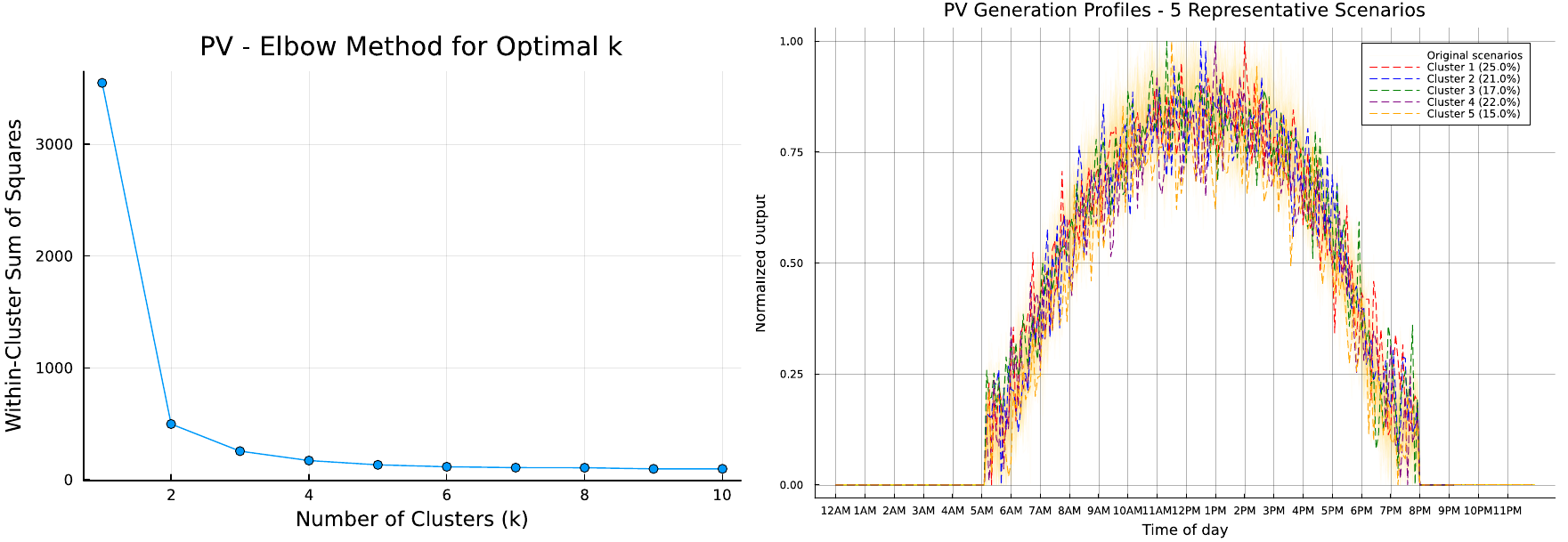}
    \caption{Scenario reduction by k-means clustering. Left: within-cluster variation stops decreasing at about 5 clusters, which sets the number of representative scenarios. Right: representative PV profiles overlaid on the sampled profiles.}
    \label{fig:scenario_reduced}
\end{figure}

As an alternative to using decomposition algorithms, we continued to use the deterministic equivalent approach, but instead used scenario reduction to accelerate the simulation. We applied k-means clustering to first cluster all the profiles in the full scenario set. This clustering allows us to identify the most representative scenarios that capture most of the trends in the original time series data. The left plot in \cref{fig:scenario_reduced} shows an example of how we applied the elbow method to identify the optimal number of clusters for k-means. We see that the within-cluster variation stops decreasing after around 5 clusters. Thus, we chose to sample 5 scenarios out of the 100 total scenarios for our analysis. The right plot shows an example of what the representative scenarios look like for the PV radiation profiles, overlaid on the raw data.

\begin{figure}
    \centering
    \includegraphics[width=\textwidth]{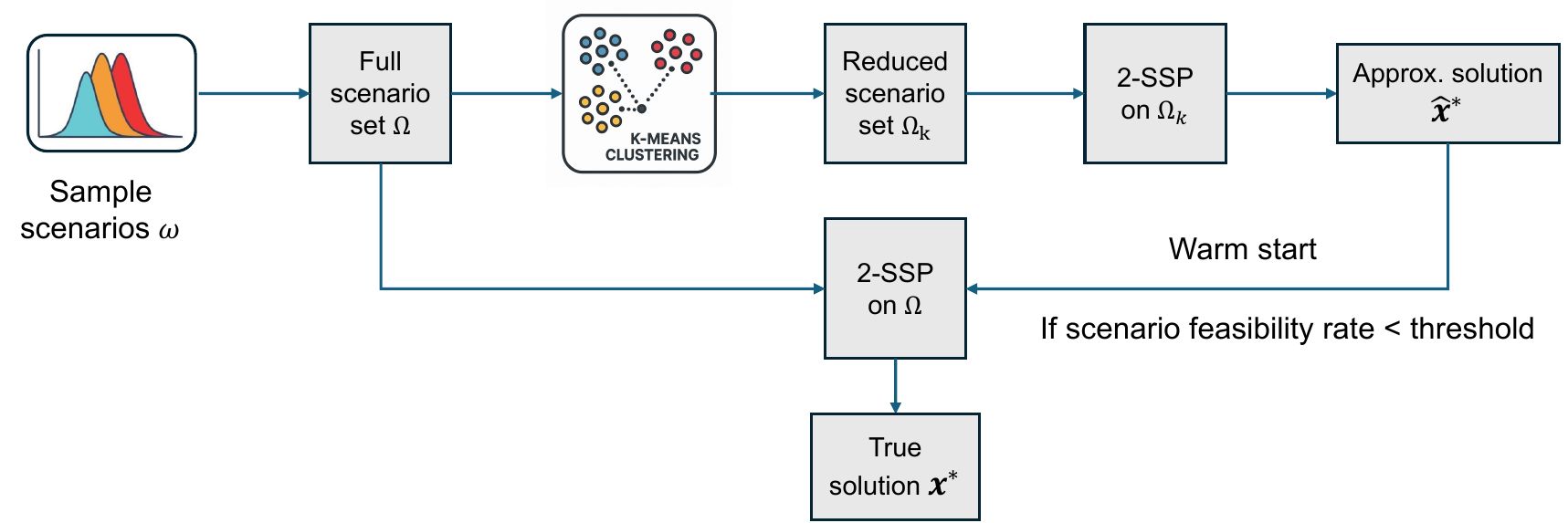}
    \caption{Accelerated 2-SSP: the problem is first solved on the reduced scenario set; if that solution's feasibility rate on the full set is below the planner's threshold, the full problem is re-solved, warm-started from the reduced solution.}
    \label{fig:accelerated_2ssp}
\end{figure}

We combined this scenario reduction with another algorithm to further accelerate the 2-SSP. This method is described in \cref{fig:accelerated_2ssp}. The problem is first solved while considering the reduced scenario set. This can be solved much more quickly due to the smaller number of scenarios. This provides an approximate solution, but it may not be feasible for all the original scenarios. In practice, though, these crude solutions can perform reasonably well in terms of feasibility. However, if the scenario feasibility rate (i.e., the percentage of original scenarios for which the approximate solution is feasible) is below the threshold set by the planner, we can re-solve the 2-SSP on the full set of scenarios. However, by using the crude solution to warm start the optimization, we can potentially accelerate convergence.

\section{Results}
\label{sec:results}

Each subsection below answers one question with one method. \cref{sec:res_pv,sec:res_hp} use the deterministic iterative method to ask how much PV and heat pump capacity coordination adds when storage and flexible loads are at low penetrations. \cref{sec:results_2ssp} uses the 2-SSP to ask how the joint PV, heat pump and storage envelope changes when siting is optimized as well. \cref{sec:res_si} uses the stochastic iterative method to ask how input uncertainty changes these gains, and \cref{sec:reconcile} reconciles the numbers from the three methods.

\subsection{Batteries and flexible demand help boost solar PV}
\label{sec:res_pv}

\begin{figure}[htbp]
    \centering
    \includegraphics[width=\textwidth]{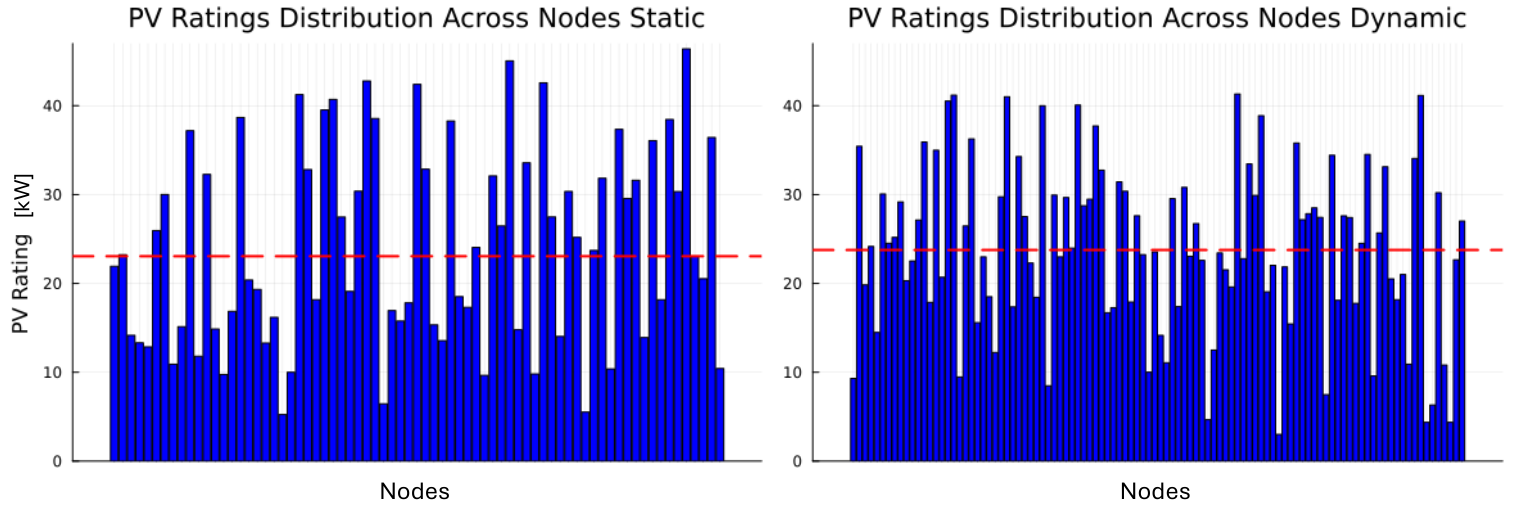}
    \caption{Distribution of nodal PV capacities at the static (48\%) and dynamic (83\%) PV hosting capacity, deterministic iterative method: coordination adds PV mainly by enabling it at more nodes, while the mean nodal capacity (red dashed line) changes little.}
    \label{fig:pv_histograms}
\end{figure}

\begin{figure}[htbp]
    \centering
    \includegraphics[width=\textwidth]{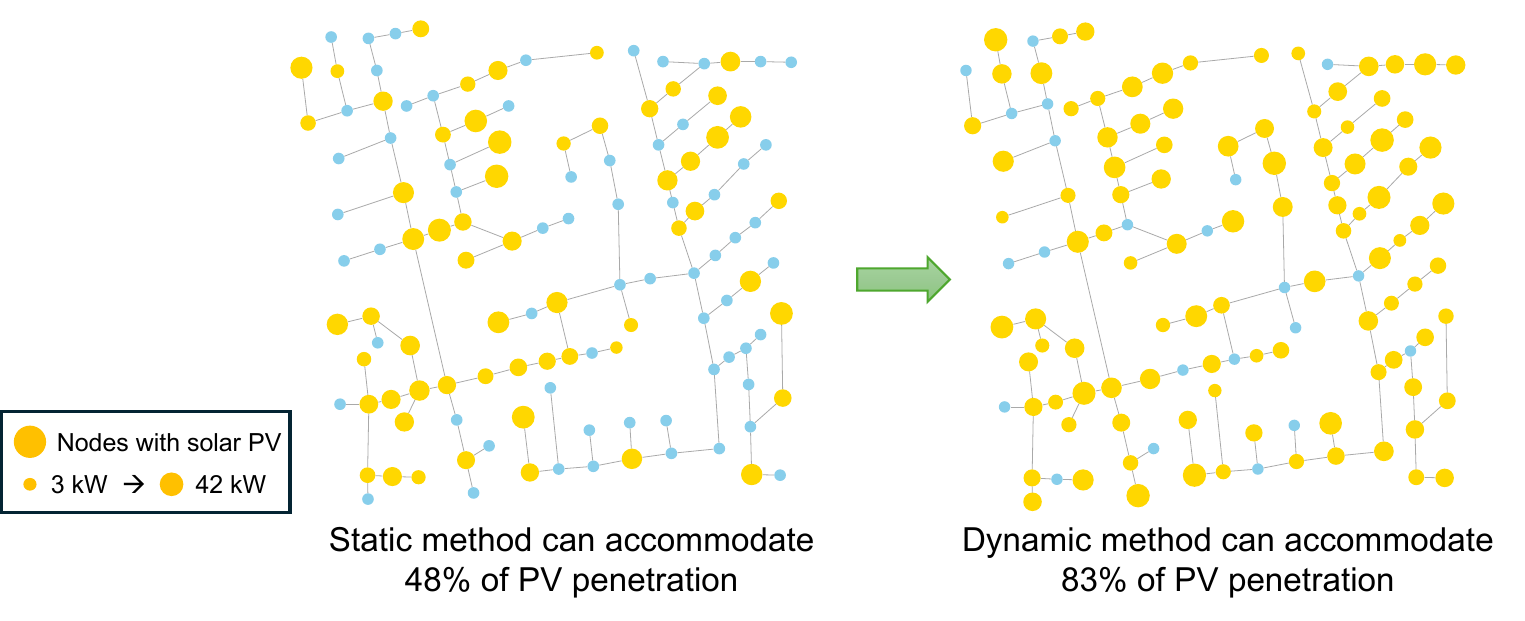}
    \caption{Coordination enables PV at new nodes as well as larger PV capacities at existing nodes (gold circles; diameter proportional to nodal PV capacity), deterministic iterative method.}
    \label{fig:pv_network_plot_det_iter}
\end{figure}

Using the deterministic iterative HCA method, we first consider the problem where we maximize PV penetration for fixed penetrations of BS, EVs, and HPs. BS penetration is set to 5\% of total feeder peak load, while 5\% of homes are assumed to have EVs and HPs. \cref{fig:pv_histograms} shows the distributions of solar capacities over the network. We see that the dynamic approach significantly increases the total PV capacity to 83\% compared to the 48\% with the static case. Thus, coordination allows for a 73\% relative increase (1.7$\times$) in solar penetration without curtailment. Although the mean capacity (shown by the red dashed line) is similar in both cases, the dynamic case has PV present at many more nodes. \cref{fig:pv_network_plot_det_iter} shows the same results but overlaid on the network plot itself to show where the PV nodes (marked by the gold circles) are located, with the diameter representing the nodal PV capacity.

\begin{figure}[htbp]
    \centering
    \includegraphics[width=\textwidth]{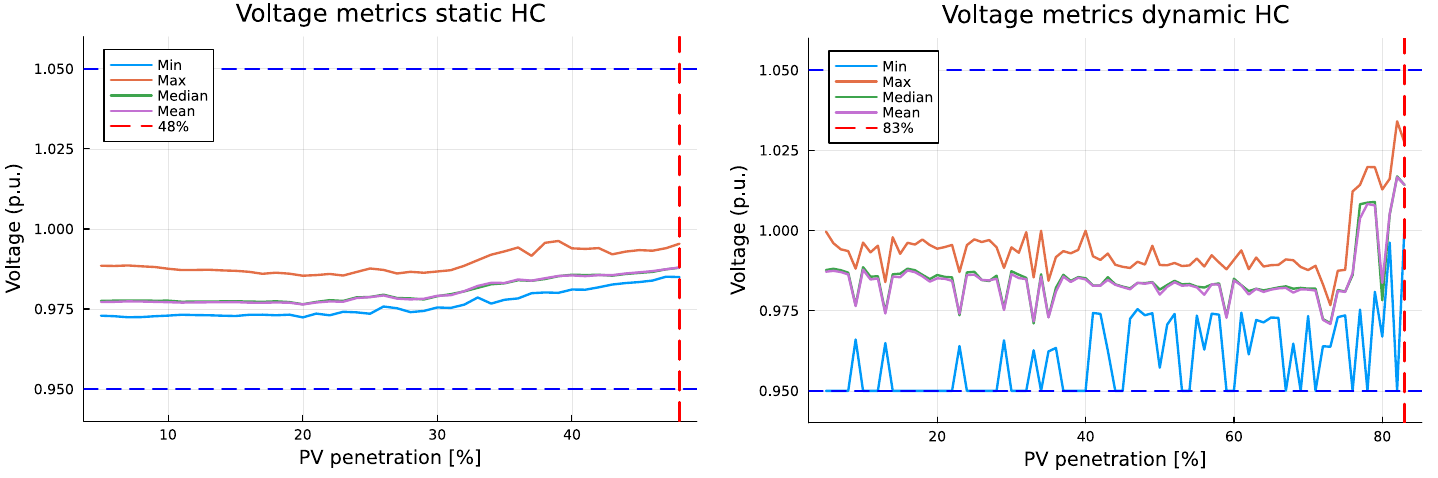}
    \caption{Voltage metrics versus PV penetration: the static case becomes infeasible above 48\% PV, while coordination keeps maximum voltage below 1.05~p.u. and mean and median voltages near 1~p.u. up to about 75\% PV, reaching its hosting capacity at 83\%.}
    \label{fig:voltage_metrics_det_iter}
\end{figure}

\begin{figure}[htbp]
    \centering
    \includegraphics[width=\textwidth]{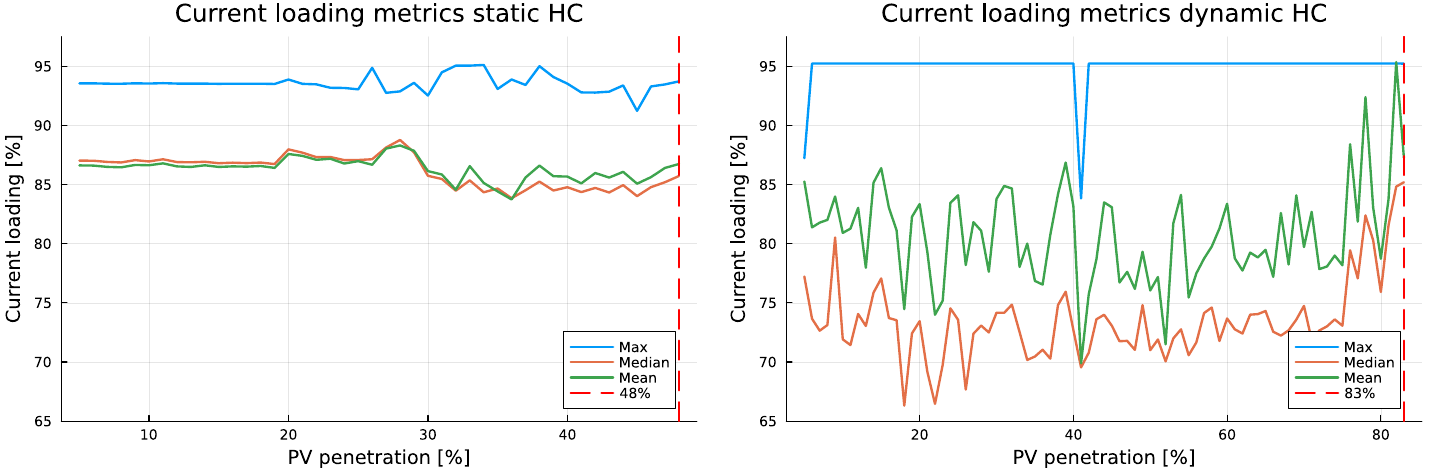}
    \caption{Line current metrics versus PV penetration: maximum currents are similar in the static and dynamic cases, but coordination substantially lowers mean and median line currents.}
    \label{fig:current_metrics_det_iter}
\end{figure}

To further interpret these results, we inspect the key power flow metrics. From \cref{fig:voltage_metrics_det_iter}, we see that the dynamic approach is able to maintain relatively low voltages up until around 75\% PV. The static case only remains feasible until 48\% PV. In addition to keeping maximum voltages well below the upper bound (1.05 p.u.), dynamic coordination also props up the mean and median voltages closer to the desired value of 1 p.u. As we increase PV beyond 75\% in the dynamic case, the voltages start to rise rapidly until we hit infeasibility at 83\% PV penetration. We also notice that there is more volatility in the voltage metrics in the dynamic case than in the static case. Comparing the two cases in \cref{fig:current_metrics_det_iter}, we find that although the maximum currents are similar, the mean and median network currents are significantly lower in the dynamic case. This indicates that leveraging flexibility with the dynamic approach has significant distributional benefits in lowering current loading throughout the network, even if the worst-case values are similar. Examining these metrics also helps us identify that nodal overvoltage and highly loaded lines are likely the main limiting factors for hosting capacity.

\begin{figure}[htbp]
    \centering
    \includegraphics[width=\textwidth]{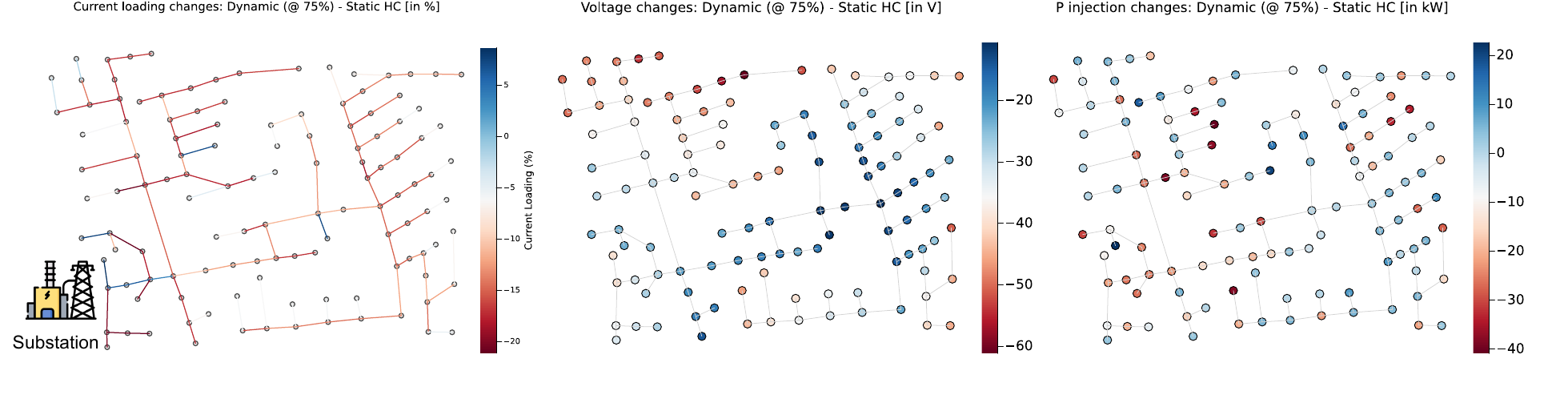}
    \caption{Changes in line currents (left), nodal voltages (middle) and nodal active power injections (right) from 48\% PV (static) to 75\% PV (dynamic) at 12:30~PM: coordination reduces overvoltage at all nodes and current loading in most lines while allowing higher injections; some lines near the substation carry more current.}
    \label{fig:metric_changes_48_75}
\end{figure}

\begin{figure}[htbp]
    \centering
    \includegraphics[width=\textwidth]{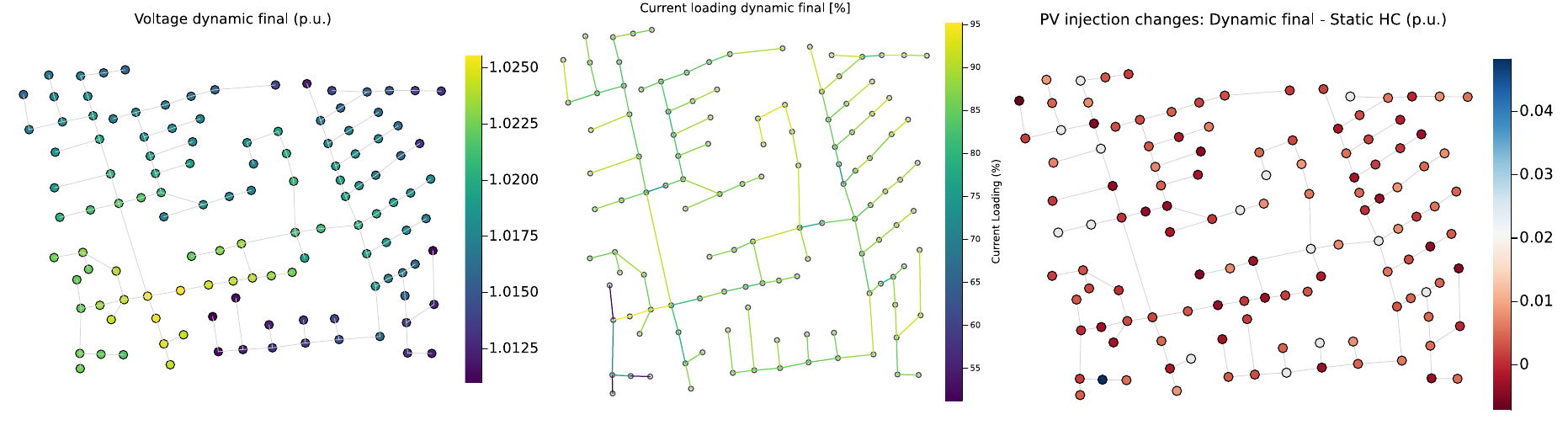}
    \caption{The network at the dynamic PV hosting capacity of 83\% at 12:30~PM: voltages and line currents remain within limits.}
    \label{fig:network_dynamic_final}
\end{figure}

\begin{figure}[htbp]
    \centering
    \includegraphics[width=\textwidth]{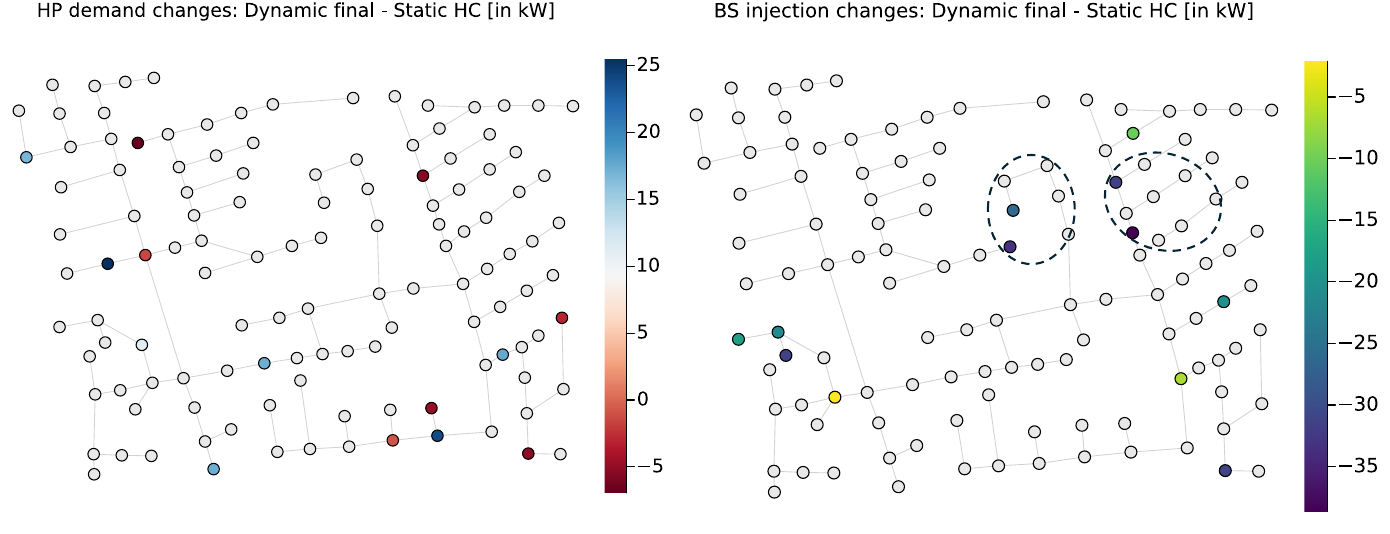}
    \caption{Heat pump and battery dispatch at 12:30~PM under coordination: heat pumps raise consumption and batteries charge (negative injection) at key nodes to absorb midday PV output.}
    \label{fig:hp_bs_boost_hc_det_iter}
\end{figure}

\cref{fig:metric_changes_48_75} shows the changes in the key metrics as we increase the PV penetration from 48\% in the static case to 75\% in the dynamic case. We see that even though we push the network to much higher PV levels, the dynamic method is able to reduce overvoltage issues at all nodes (middle plot) and quite significantly bring down current loading in most lines (left plot), although there are increases in some lines, particularly near the substation where the distribution grid is connected to transmission. This is a good sign as it indicates that the network is able to handle higher levels of PV without overloading or overvoltage issues, allowing higher active power injections at many nodes in the network (as seen in the right plot). Note that \cref{fig:metric_changes_48_75,fig:network_dynamic_final,fig:hp_bs_boost_hc_det_iter} are all network plots for a single snapshot of time during midday (12:30~PM local time). This time instant is of special interest to us since it corresponds to the peak PV output, which is most likely to cause overvoltage and current load issues.

In \cref{fig:network_dynamic_final}, we show the final state of the network at the maximum dynamic PV HC of 83\%, and we see that even as we push PV injections to the maximum feasible level, voltages and currents remain within limits at every simulated time step.

\begin{figure}[htbp]
    \centering
    \includegraphics[width=0.6\textwidth]{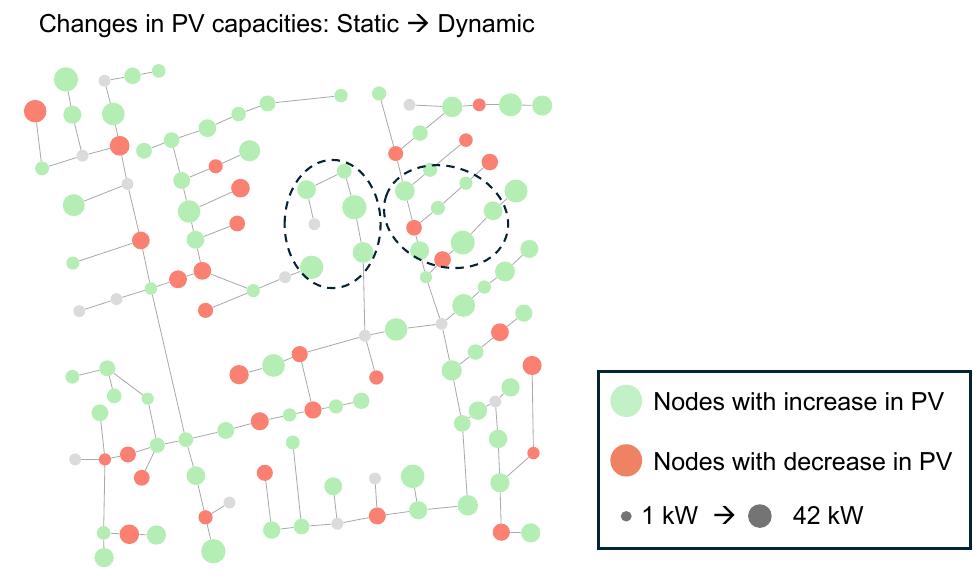}
    \caption{Change in nodal PV capacity from the static (48\%) to the dynamic (83\%) PV hosting capacity, deterministic iterative method.}
    \label{fig:pv_size_changes_network_det_iter}
\end{figure}

We next examine how network-aware coordination supports increased PV hosting capacity. \cref{fig:pv_size_changes_network_det_iter} shows the changes in the PV capacities at nodes between the dynamic case (83\%) and static case (48\%). Comparing this against the BS and HP dispatch plots in \cref{fig:hp_bs_boost_hc_det_iter}, we see that the flexible batteries and heat pumps play a key role in supporting PV. Note that positive BS injections represent discharging while negative injections indicate charging. Essentially, the heat pumps consume more power at several key nodes, and the batteries charge more in order to help absorb the excess PV output midday. Heat pump consumption can be increased by adjusting indoor temperature setpoints to be lower, thus raising cooling demand. The total BS power is 147~kW (5\% of the total peak load of 2,930~kW) with an energy storage capacity of 441~kWh, while heat pumps in 5\% of homes (about 42 homes) add about 0.2~MW of flexible load. The network also has EVs in 5\% of homes, but these are only available for limited periods. Many residential EVs are not available during the middle of the day when they are likely away from home, and thus they do not play as important a role here. EV charging will likely be more helpful to reduce demand overnight when charging usually occurs. Comparing the clusters (dashed circles), we see that co-located batteries help boost PV at the same node, but batteries also support PV at nearby nodes due to the network effect.

\subsection{Coordination helps accommodate load growth from heating and transportation electrification}
\label{sec:res_hp}

\begin{figure}[htbp]
    \centering
    \includegraphics[width=\textwidth]{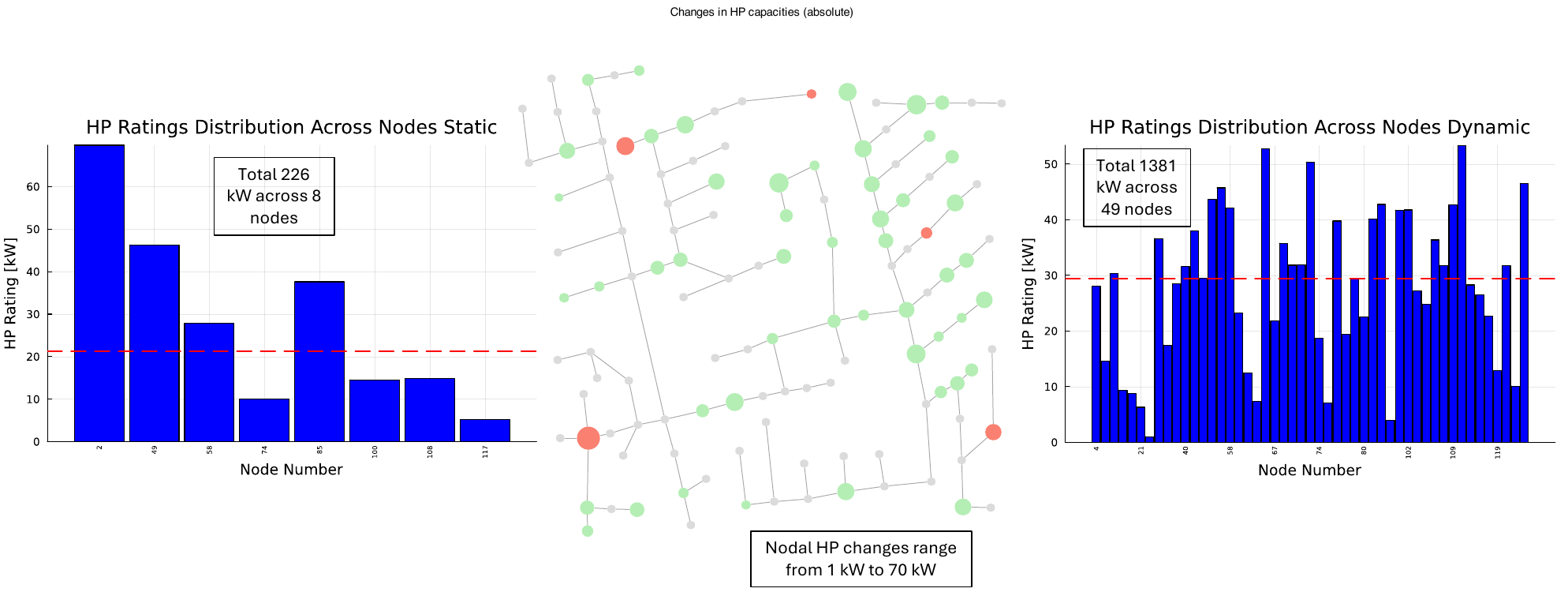}
    \caption{Distribution of nodal heat pump penetrations at the static (9\%) and dynamic (55\%) heat pump hosting capacity, with 10\% storage, 20\% PV and 5\% EVs, deterministic iterative method.}
    \label{fig:hp_max_histograms}
\end{figure}

Using the deterministic iterative HCA method again, we show a case study on how we can boost HP hosting capacity by leveraging all the available DERs. This is crucial since heating and cooling generally contribute the most to residential home load, after EVs (if present). The transition away from gas-based heating and conventional air conditioning systems towards electric heat pumps will result in large demand increases, which can stress the distribution grid. The following results consider maximizing the HP penetration on a system with 10\% BS penetration, 20\% PV penetration, and 5\% of homes with EVs. By using dynamic coordination, we are able to increase the HP penetration from 9\% in the static case to 55\% in the dynamic case, as seen in \cref{fig:hp_max_histograms}.

\begin{figure}[htbp]
    \centering
    \includegraphics[width=\textwidth]{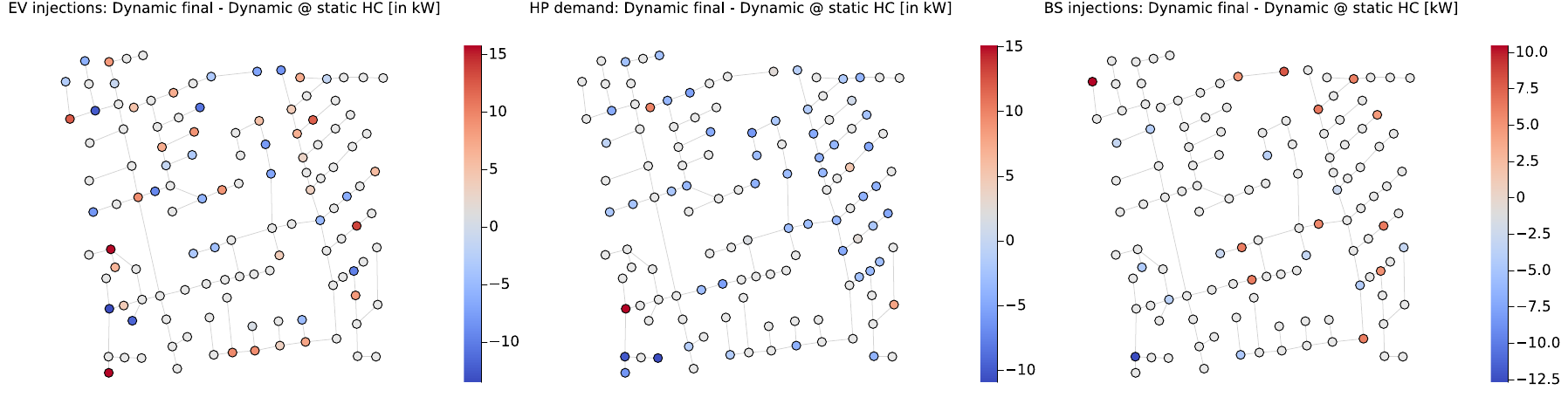}
    \caption{Changes in DER injections at 7:30~PM as heat pump penetration rises from 9\% to 55\% under coordination: heat pump demand is reduced through setpoint adjustment at most nodes and battery and EV discharging increases, lowering net load.}
    \label{fig:hp_max_der_changes}
\end{figure}

To illustrate how dynamic coordination helps boost HP HC, we zoom in to focus on one hour (7-8 PM) of evening operation since this period represents peak load stress on the grid due to the loss of solar PV output along with large ramps in HP and EV consumption as well as other inflexible loads. The added HP load could correspond to either heating or cooling, depending on the season. This is a completely different setting from the period of 12-1 PM that we focused on in the PV maximization case, which corresponded to excess generation. \cref{fig:hp_max_der_changes} shows a snapshot of the network at 7:30 PM, with the changes in power injections from various DERs. We compare the injections between two different HP penetration levels, both under the dynamic case, to see how the DER injections change as we increase HP levels from 9\% (the maximum static HP HC) to 55\% (maximum dynamic HC). Here, we see that the dynamic approach performs significant curtailment of flexible HP demand by adjusting temperature setpoints accordingly. However, there are a few exceptions since the HP load does increase at a small number of nodes, mainly driven by the increase in their baseline installed capacities at higher penetrations. In addition, we generally see an increase in discharging of both BS and EVs throughout the system. These changes bring down the net load and thus reduce stress on the grid. Note that the EV charging demand does increase at several nodes. This is because EV charging demand generally peaks in the evening and overnight after cars return home, making it relatively less elastic. Our optimization framework ensures that the changes resulting from dynamic load shifting (or curtailment) and BS dispatch still respect all DER constraints (e.g., state of charge, interior temperature), as well as customer preferences (e.g., thermal comfort, cycling penalties) through the objective function.

\begin{figure}[htbp]
    \centering
    \includegraphics[width=\textwidth]{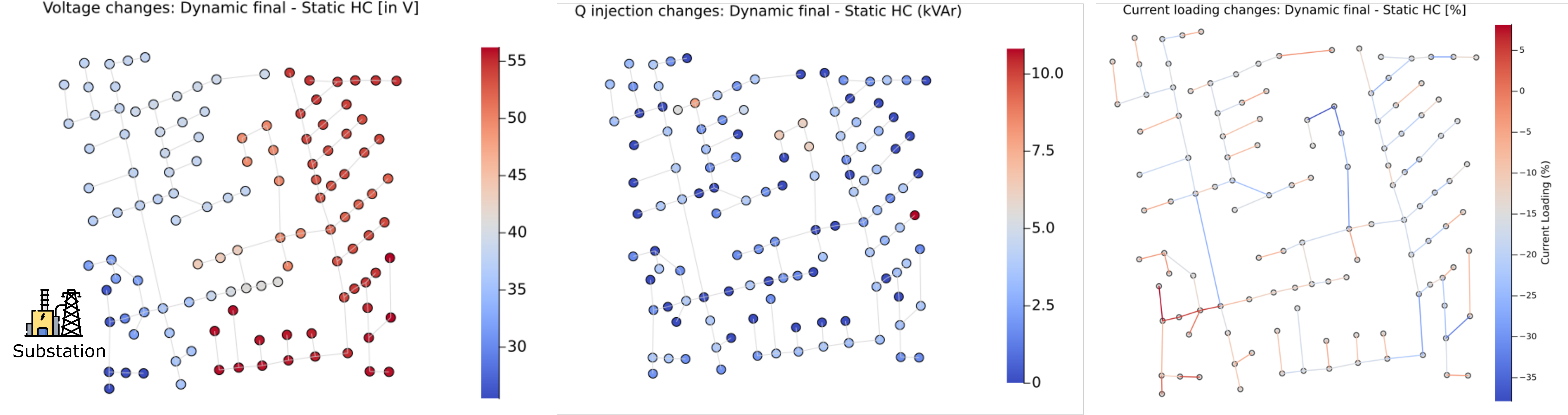}
    \caption{Changes in voltages (left), inverter reactive power (middle) and line currents (right) at 7:30~PM between the static (9\%) and dynamic (55\%) heat pump hosting capacity: coordination raises voltages most at nodes deep in the feeder, increases reactive support, and lowers currents in most lines.}
    \label{fig:hp_max_VIQ_changes}
\end{figure}

The effects of dynamic coordination on the key power flow metrics are shown in \cref{fig:hp_max_VIQ_changes}, where we compare the network at the maximum static versus dynamic HC. On the leftmost plot, we see that dynamic coordination boosts the voltages throughout the network compared to the static case, to counter the voltage drops due to increased HP load. The voltage rise increases as we move deeper into the feeder away from the substation since these nodes experience more severe undervoltage issues, which follows from Ohm's law and increased line losses. The undervoltage risks are relatively lower closer to the substation since the distribution feeder imports more power from the main transmission grid. The middle plot shows that the dynamic approach increases reactive power injections from inverters throughout the network, which are crucial to provide voltage support. The rightmost plot shows that there is a general decrease in current loading throughout the network since dynamic coordination relieves thermal constraints via DER flexibility. However, there is an increase in congestion in a few lines mainly near the substation due to an increase in imported power.

\subsection{Dynamic coordination enables strong complementary relationships among different types of DERs}
\label{sec:results_2ssp}

\begin{figure}[htbp]
    \centering
    \includegraphics[width=\textwidth]{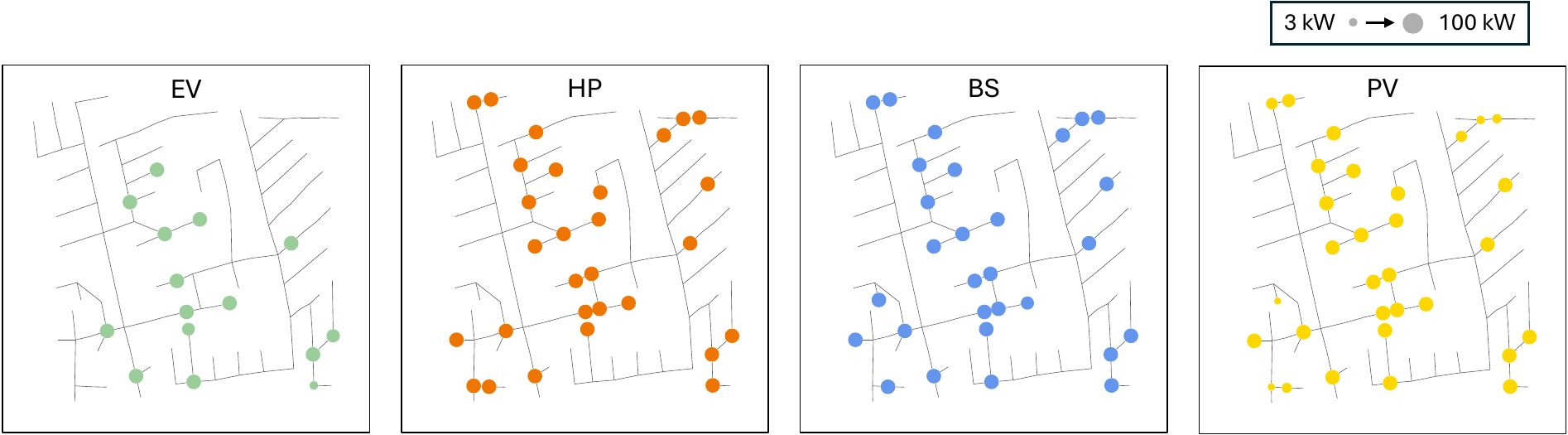}
    \caption{DER siting and sizing from the dynamic 2-SSP (20\% storage, 20\% heat pumps, 10\% EVs, 5 scenarios): capacities are spread evenly across nodes and storage tends to be co-located with PV; PV hosting capacity reaches 98\%.}
    \label{fig:der_dist_2ssp}
\end{figure}

Here we present selected results obtained by solving the two-stage stochastic program (2-SSP) on the IEEE 123-node test network with 20\% BS penetration, 20\% of homes equipped with HPs, 10\% of homes with EVs, and by considering $K = 5$ representative scenarios identified by scenario reduction. \cref{fig:der_dist_2ssp} shows the dynamically optimized locations and capacities of each of the four DER types. Interestingly, we see that the optimization solution results in a generally uniform and balanced distribution of capacities over nodes and also favors co-locating batteries with solar. Electrified loads (HP, EV) also tend to be co-located with batteries and/or solar. Note that this co-location was not enforced by design but is rather just an outcome of the optimization. An intuitive explanation for this phenomenon could be that co-location enables each node to be more self-sufficient (by using self-generation to meet self-demand), and thus reduces its net injections into the grid (or power drawn by it from the grid). This reduces the stress on the grid resulting from additional PV or demand and thus allows the network to host more DERs. The total dynamic PV HC for this case was determined to be 98\%, compared to a static PV HC of 24.2\%.

\begin{figure}[htbp]
    \centering
    \includegraphics[width=0.5\textwidth]{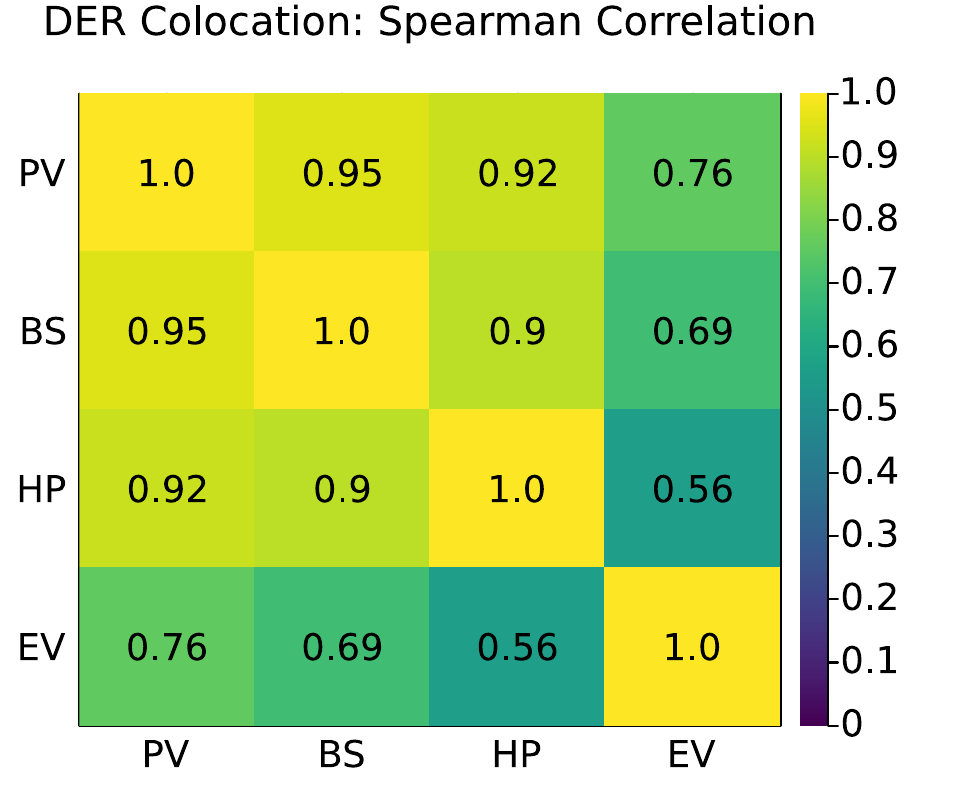}
    \caption{Spearman rank correlations between co-located nodal capacities of each DER pair in the dynamic 2-SSP solution. All are positive, partly because every DER size scales with nodal load or the number of homes at a node.}
    \label{fig:der_coloc_corr}
\end{figure}

\cref{fig:der_coloc_corr} shows the Spearman correlations of co-located capacities of different types of DERs at the same node, while using dynamic coordination. All the pairwise correlations are positive, indicating a tendency for DER types to be co-located under dynamic coordination. Two caveats apply. First, PV and storage are both sized relative to nodal load, and heat pumps and EVs relative to the number of homes at a node, so larger nodes tend to receive more of every DER type; part of each positive correlation therefore reflects node size rather than complementarity. Second, the MISOCP optimum is not unique, so other siting patterns with the same total penetration may exist. With these caveats, the pattern is consistent with co-location helping DERs support one another in maintaining feasibility. We see that PV is very strongly correlated with both BS and HP. This is plausible because excess PV output during the day can be utilized for both BS charging and to meet heating or cooling demand. The correlation between PV and EV is weaker since EV charging mainly dominates during time periods when solar PV generation is not available. BS is also strongly correlated with HPs since BS discharging can help support HP load during peak demand periods. However, BS and EVs are less strongly correlated since they both essentially function as energy storage devices and their charging and discharging profiles may not always align in a complementary fashion. Finally, we see that the weakest correlations are between HP and EV since these are both loads and may compete for power consumption during some periods, especially overnight. Their correlation is nonetheless positive, which is consistent with their being able to support each other, e.g., the HP can shift or curtail its demand to make more room for EV demand, or vice versa.

\begin{figure}[htbp]
    \centering
    \includegraphics[width=\textwidth]{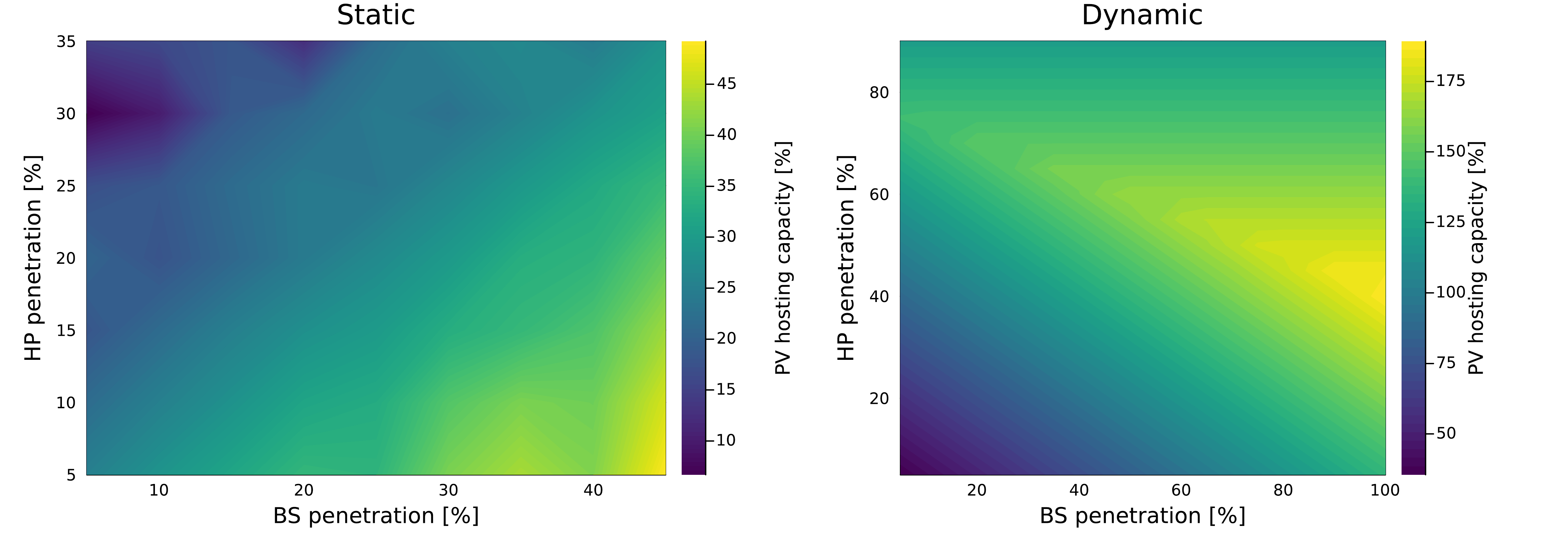}
    \caption{Maximum PV penetration over feasible storage and heat pump penetrations, static (left) and dynamic (right) 2-SSP: under static operation storage and heat pumps act as weak substitutes, while coordination makes them complements and enlarges the feasible region about 22-fold.}
    \label{fig:2ssp_sensitivity}
\end{figure}

The accelerated version of the 2-SSP with scenario reduction and warm start also enables rapid sensitivity studies. \cref{fig:2ssp_sensitivity} shows the results for a sensitivity case study where we varied both the BS and HP penetrations in order to assess their effects on the maximum possible PV penetration attainable. For each case, we only plotted the combinations of (BS, HP, PV) penetrations that are power flow-feasible. The left plot shows results for the static case. The direction of the contour lines suggests that BS and HP penetration have opposing or competing effects on PV hosting capacity, due to limiting grid constraints and uncoordinated, sub-optimal operation. Increasing BS penetration does help boost PV but increasing HP tends to reduce PV capacity. In addition, more of the variation occurs along the x-axis, indicating that BS has a larger impact than HPs when there is no coordination. In addition, we see that as we move towards the upper left corner, the HP penetration starts to become larger than BS, resulting in lower PV penetrations. On the other hand, points towards the lower right-hand corner with higher BS penetrations relative to HP allow higher PV penetrations. Thus, under the static analysis, the HP and BS have a weak substitutive relationship where BS is the primary driver of PV HC gains while higher HP penetration tends to reduce the utility of BS and erodes the available headroom to increase PV. This makes it challenging to increase penetrations of all three DERs simultaneously. The feasible region is thus limited, only allowing up to about 45\% of BS and 35\% of maximum HP penetration, with a maximum achievable PV penetration of roughly 50\%.

The right plot shows the results for the dynamic case. The contour lines suggest that BS and HP have complementary effects on PV hosting capacity since higher levels of both BS and HP penetration together generally lead to higher PV hosting capacity. This is because dynamic coordination allows high BS levels to compensate for high HP and maintain high PV levels. However, when we approach high HP penetrations over 50\%, this complementary relationship weakens. While increasing BS beyond this point continues to increase PV, increasing HP further starts to slightly degrade PV levels. However, we are still able to maintain very high PV penetrations above 100\% even after this stage, as we approach the upper right corner of the graph (which represents nearly full electrification of both heating and transport). By relieving grid constraints in this manner, we can significantly expand the feasible space: with storage up to 100\% of peak load, the feeder hosts heat pumps in up to 90\% of homes and PV up to around 200\% of peak load. The smooth gradient indicates that both technologies have complementary effects. We see that the dynamic approach can accommodate much higher penetrations of all three DERs. We also see that there is more variation along the BS axis. Tracing horizontally across any given HP penetration level, the PV capacity shifts dramatically from deep blue (25\%) at low BS penetration to bright yellow (175\%+) at high BS penetration. In contrast, tracing vertically at a fixed BS penetration, the color change is much more muted, especially at low-to-moderate BS levels where the plot stays relatively uniform in the teal-green range. This indicates that while both resources can help absorb and buffer excess PV output, storage has a stronger complementary effect with PV than HPs. We focus specifically on BS, HP, and PV in these sensitivity studies because the distinct temporal signatures in their profiles allow for strong complementary effects. EVs are more challenging to leverage and model in this manner due to added complexities in their availability, in terms of whether they are charger-connected at specific times or have bidirectional charging enabled, as well as greater concerns around cycling, lifetime and warranty. Consequently, these results do not rank EVs against storage and heat pumps. \cref{fig:2ssp_sensitivity} also shows that the dynamic approach leads to a roughly 22-fold increase (over the static case) in the feasible combinations of DER penetrations, as measured by the volume of the feasible (PV x HP x BS) penetration levels.

\begin{figure}[htbp]
    \centering
    \includegraphics[width=\textwidth]{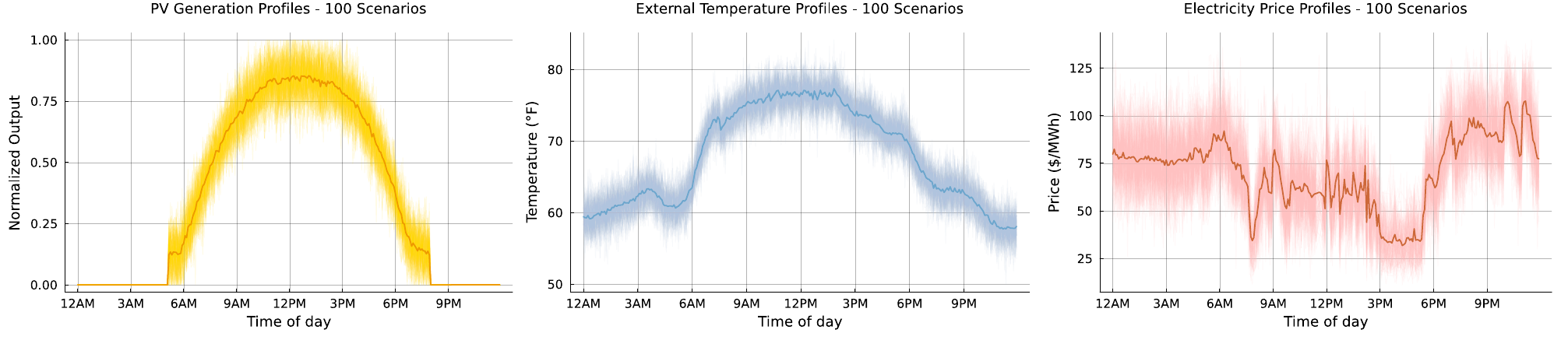}
    \caption{Key input profiles: midday PV output coincides with high outdoor temperature (cooling demand) and low LMPs, which underlies the complementarity of PV with heat pumps and storage.}
    \label{fig:der_complementarity}
\end{figure}

\cref{fig:der_complementarity} clearly shows the complementary relationships among different DERs based on the key inputs. We see that high PV output in the middle of the day also coincides with higher external temperatures and greater cooling demand, thereby allowing HPs to contribute their flexibility. During winter months, we expect EVs to play more of a role since their peak demand overnight also complements lower temperatures and higher heating demand, as well as peak wind power production. Similarly, electricity prices are also lower mid-day (mainly due to excess PV output), which further incentivizes batteries to charge more when cheap power is available.

\begin{figure}[htbp]
    \centering
    \begin{subfigure}[b]{0.49\linewidth}
    \includegraphics[width=\linewidth]{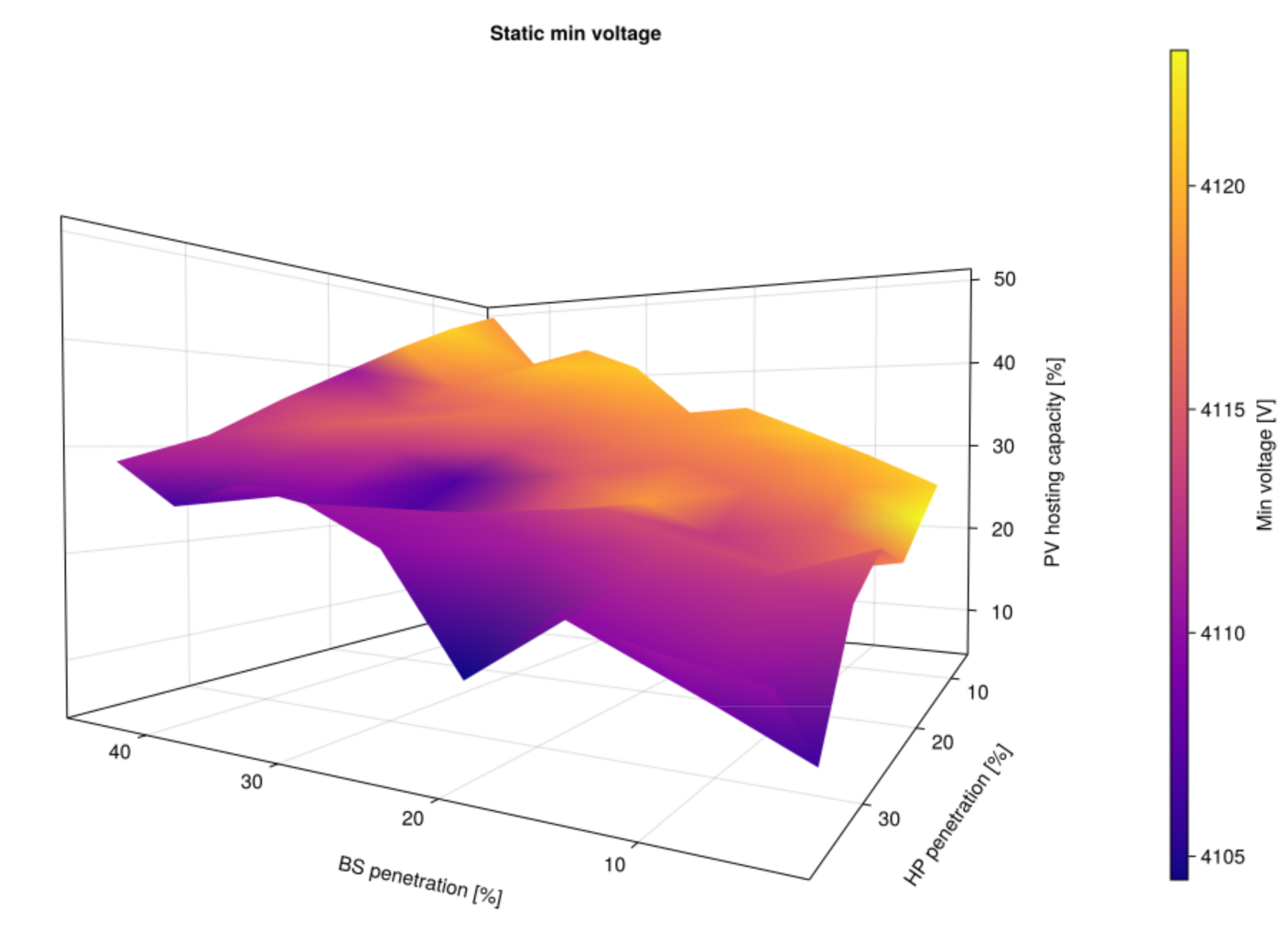}
    \caption{Static case minimum voltage.\label{fig:minV_compare_2ssp1}}
    \end{subfigure}
    \begin{subfigure}[b]{0.49\linewidth}
    \includegraphics[width=\linewidth]{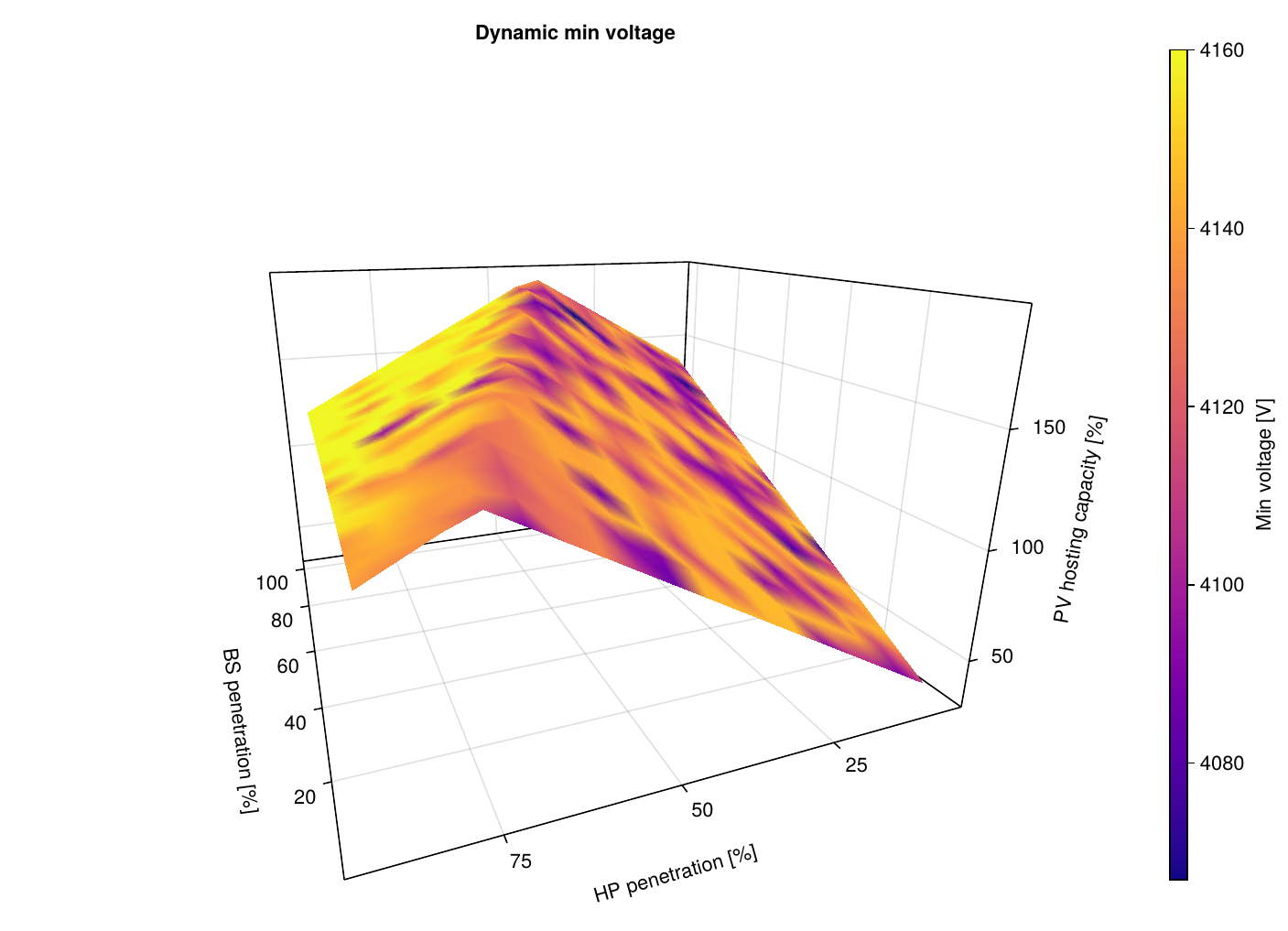}
    \caption{Dynamic case minimum voltage.\label{fig:minV_compare_2ssp2}}
    \end{subfigure}
        \caption{Minimum network voltage over feasible heat pump, storage and PV penetrations from the 2-SSP: storage has little effect on minimum voltage under static operation, while under coordination minimum voltages stay healthy across most of the feasible space.\label{fig:minV_compare_2ssp}}
\end{figure}

To gain further insights into the effects of varying different DER penetrations on power flow, we can look at the sensitivity of voltage and current metrics to such changes. These were either the minimum or maximum values observed across all the considered scenarios and over the entire network and simulation period. \cref{fig:minV_compare_2ssp} compares the minimum recorded voltages between the static versus dynamic cases. This undervoltage risk is mainly driven by HP load and line losses. Note that the nominal base voltage here is 4160 V, so any values below or above that correspond to under- or over-voltage conditions, respectively. In the static case, as expected, we see that increasing PV penetration reduces undervoltage issues since the additional generation props up voltages. Similarly, increasing HP levels makes undervoltage issues worse since without coordination, their inflexible load just brings down voltages. Interestingly, we also see that across different HP and PV penetrations, increasing BS generally seems to have little to no impact on voltages. In fact, uncoordinated charging and discharging of batteries can actually make issues like reverse power flows and load swings worse, thereby reducing effective static grid hosting capacity. This further stresses the importance of network-aware optimization and coordination methods to fully extract value from additional flexible resources. This supports our claim that coordination is crucial to truly capture the benefits of battery storage. In the dynamic case, we notice that the trends are much more non-uniform and non-monotone. The surface is predominantly yellow or orange, indicating that minimum voltages remain quite healthy across the board. The dark purple voltage drops tend to occur at points with moderate HP penetration combined with low PV penetration, i.e., to the right of the ridge. Unlike the overvoltage case (discussed below) with very strong direct complementarity between BS and PV, undervoltage issues are influenced by a combined interaction of HP load flexibility, BS operation, and PV availability. Reactive power injections from smart inverters of both PV and BS can prop up voltages. Interestingly, we see that combining high HP with high BS penetrations enables more PV (to the left of the ridge), which actually mitigates voltage drops. This also means that high BS is crucial to make high PV penetration feasible.

\begin{figure}[htbp]
    \centering
    \begin{subfigure}[b]{0.49\linewidth}
    \includegraphics[width=\linewidth]{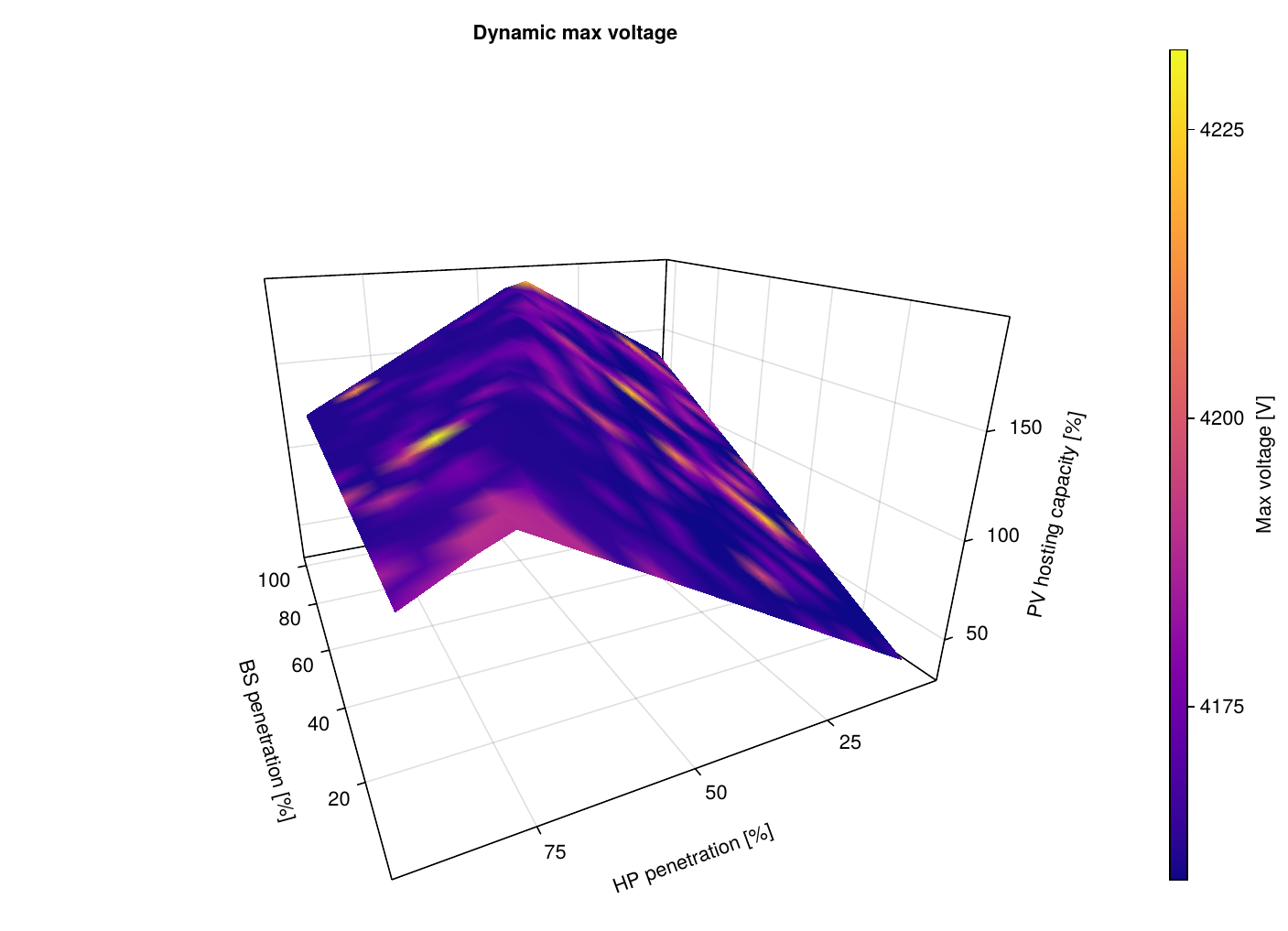}
    \caption{Dynamic case maximum voltage.\label{fig:maxVI_dynamic_2ssp1}}
    \end{subfigure}
    \begin{subfigure}[b]{0.49\linewidth}
    \includegraphics[width=\linewidth]{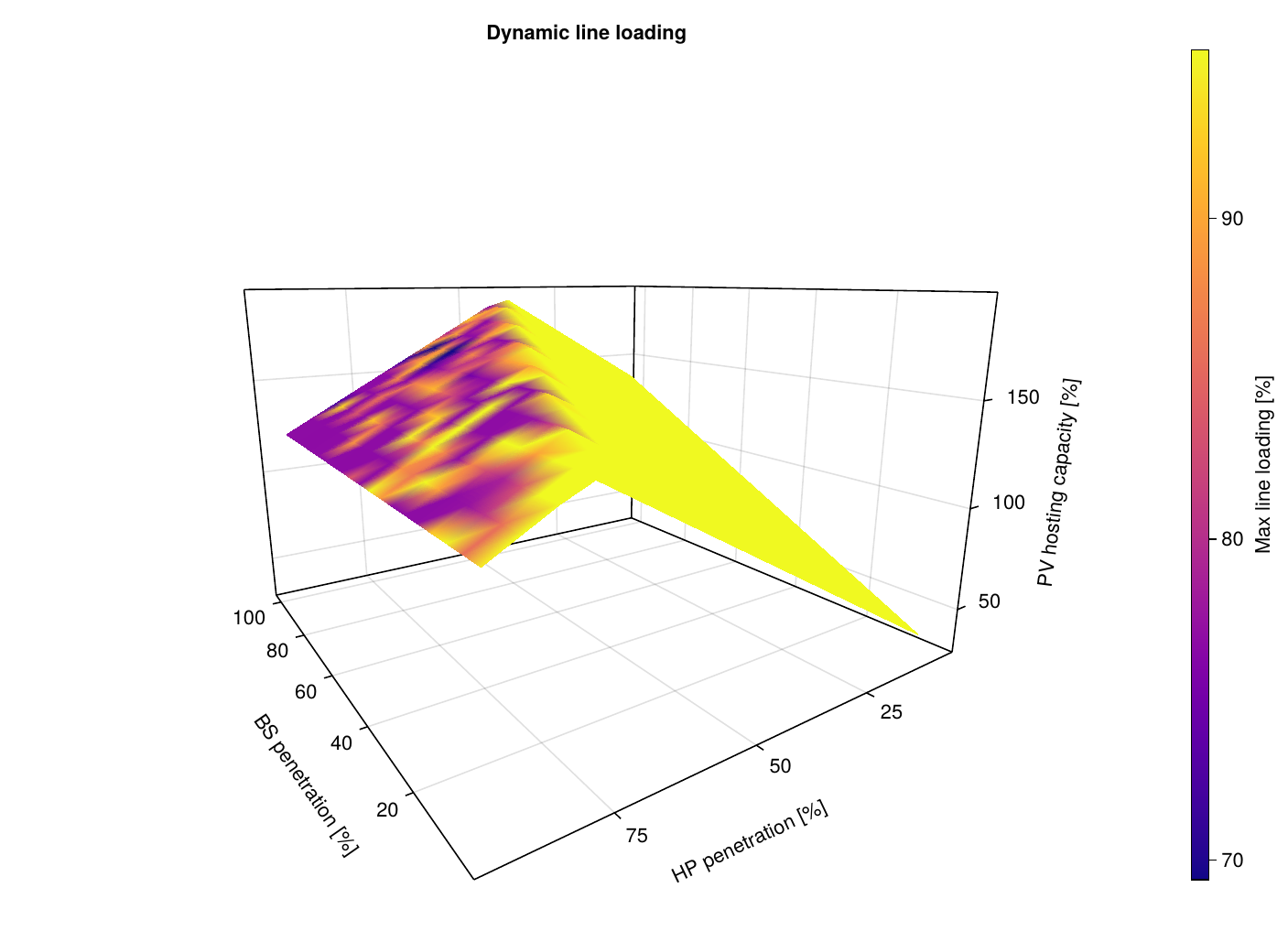}
    \caption{Dynamic case maximum current loading.\label{fig:maxVI_dynamic_2ssp2}}
    \end{subfigure}
    \caption{Maximum network voltage (a) and maximum line loading (b) over feasible penetrations, dynamic 2-SSP: maximum voltage stays below the 1.05~p.u. limit across the feasible space, while line loading comes closest to its limit and is relieved mainly by heat pump flexibility.\label{fig:maxVI_dynamic_2ssp}}
\end{figure}

\cref{fig:maxVI_dynamic_2ssp} shows the variations in maximum voltage and maximum current loading over various DER penetrations. From \cref{fig:maxVI_dynamic_2ssp1}, we see that the surface is predominantly deep blue/purple which indicates that maximum voltages stay well below the overvoltage threshold (1.05 p.u. or 4370 V for a 4.16 kV system) across the entire feasible space. The lighter (pink or yellow) regions generally occur at cases with high PV and low-to-moderate HP penetration, which makes sense since PV (coupled with low load) drives voltage rise issues due to reverse power flows during midday solar peaks. This trend holds across most BS levels. We also see that higher BS penetrations generally darken the surface more since coordinated charging offers effective down regulation of voltage. However, there is a trade-off since this also allows for higher PV penetrations, which has an opposite effect on voltages. The fold or ridge in the surface plot (in all the dynamic case plots in \cref{fig:minV_compare_2ssp2} and \cref{fig:maxVI_dynamic_2ssp}) is due to the relationships between the feasible penetration levels of the three DERs, which is also observed in \cref{fig:2ssp_sensitivity}.

\cref{fig:maxVI_dynamic_2ssp2} shows the maximum current loading as a percentage of the thermal ampacity limits of lines, which is a measure of network congestion. Of all the metrics, this has the starkest variation, especially along the HP axis. Increasing HP load generally increases the demand for power imports from the transmission grid and also increases line flows in the network, but BS charging, PV output, and load shifting or curtailment (primarily through HP thermal flexibility) can help compensate and mitigate this. The bright yellow plateau to the right of the ridge represents near-limit or binding thermal constraints and is primarily driven by low-to-moderate HP penetration, which reduces the capability of demand flexibility to mitigate current issues. At high HP penetration levels (to the left of the ridge), we see that current loading is relatively lower and further decreases to darker colors as we increase BS penetration. This is despite the high PV levels, since both BS and HP can help absorb it and reduce reverse power flows. These results indicate that HP flexibility plays the primary role in relieving thermal constraints, followed by BS and PV playing a secondary role.

\subsection{Input uncertainty significantly affects hosting capacity}
\label{sec:res_si}

\begin{figure}
    \centering
    \includegraphics[width=\textwidth]{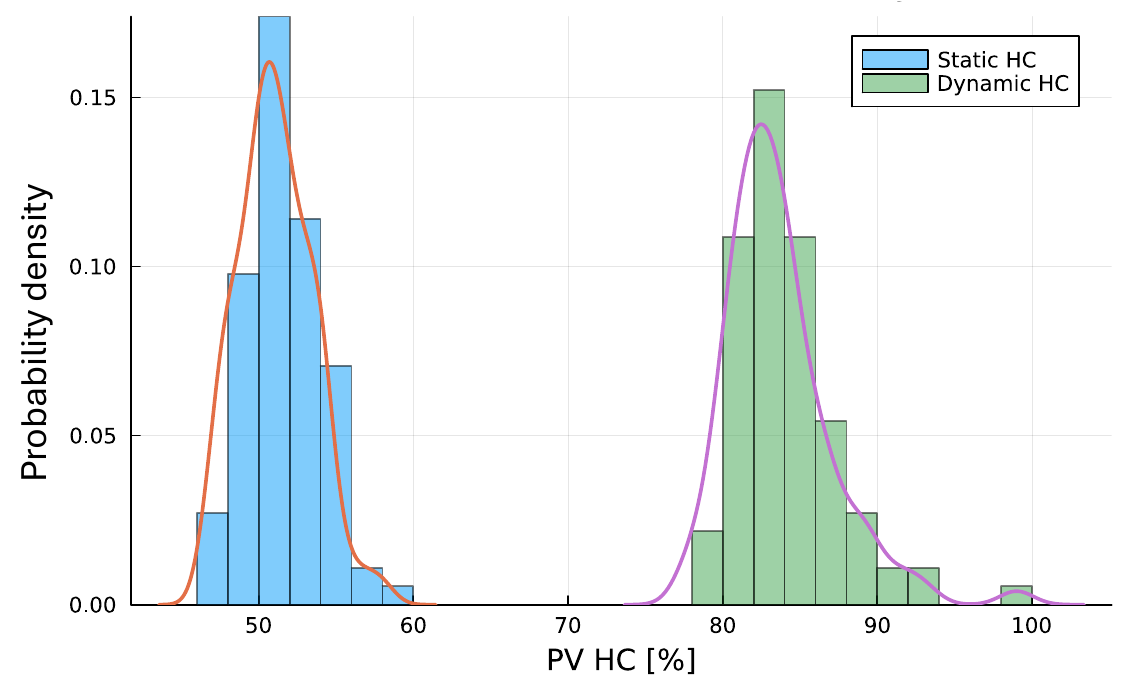}
    \caption{Distributions of PV hosting capacity over 100 sampled scenarios, with kernel density estimates: coordination shifts the whole distribution to higher values (mean 51\% static, 84\% dynamic).}
    \label{fig:stoch_iter_hc_dist}
\end{figure}

We now discuss the results of the stochastic iterative approach, where we maximized PV capacity with 5\% penetrations of both BS and HP. \cref{fig:stoch_iter_hc_dist} shows the distributions of the PV HC estimates obtained by running the method for 100 randomly sampled scenarios. Similar to the deterministic case, we see that the dynamic approach pushes the entire distribution further towards the right with higher HC values. As \cref{tab:stoch_iter_hc_dist} shows, coordination raises mean PV HC by 64\%, from 51.1\% to 83.6\%. The absolute spread also widens: the standard deviation rises from 2.36 to 3.44 percentage points. Relative to the higher mean, however, the spread narrows slightly, from a coefficient of variation of 4.6\% to 4.1\%. The wider absolute spread is likely because input uncertainty affects the DER flexibilities and thus the constraints of the optimization problem solved in the dynamic case; uncertainty that propagates through optimization constraints can be amplified \cite{bertsimas2004price}. For planning, this means coordinated HC should be sized to a low percentile of its distribution, or to the worst case across scenarios as done here, rather than to its mean.

\begin{table}
    \centering
    \begin{tabular}{@{}lcc@{}}
    \toprule
         & Static & Dynamic \\
    \midrule
         Mean HC [\%] & 51.08 & 83.63 \\
         Standard deviation [percentage points] & 2.36 & 3.44 \\
         Coefficient of variation [\%] & 4.62 & 4.11 \\
    \bottomrule
    \end{tabular}
    \caption{Summary statistics of the PV HC distributions over 100 scenarios, stochastic iterative method. Coordination raises the mean by 64\% and the standard deviation by 46\%, so the spread relative to the mean narrows slightly.}
    \label{tab:stoch_iter_hc_dist}
\end{table}

\begin{figure}[htbp]
    \centering
    \includegraphics[width=\textwidth]{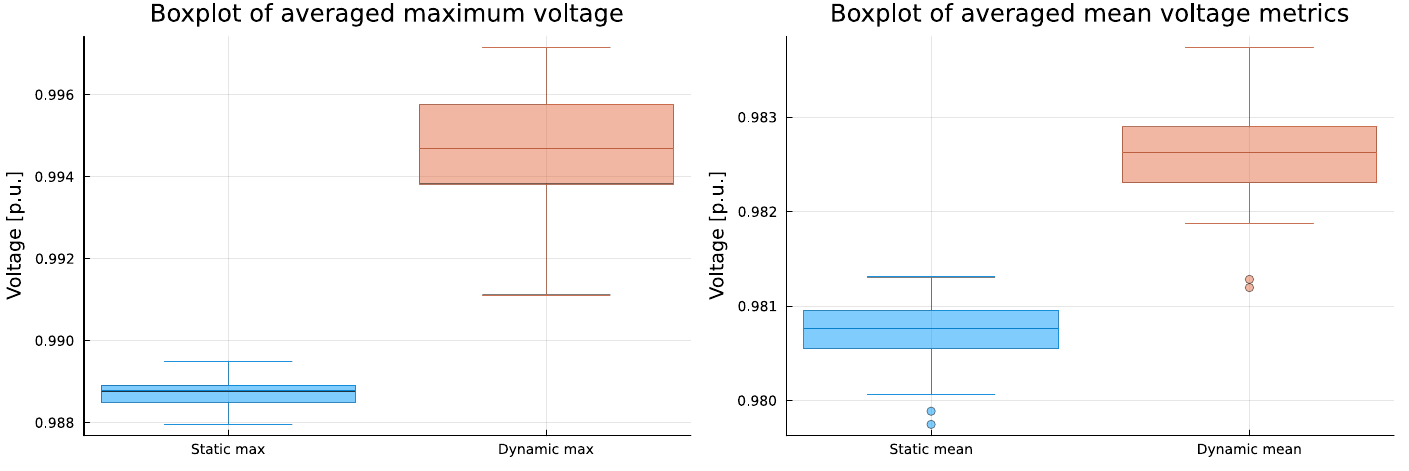}
    \caption{Distributions of maximum (left) and mean (right) network voltage over 100 scenarios, stochastic iterative method: both are higher under coordination, which hosts more PV, and the maximum voltage varies more.}
    \label{fig:stoch_iter_v_dist}
\end{figure}

\begin{figure}[htbp]
    \centering
    \includegraphics[width=\textwidth]{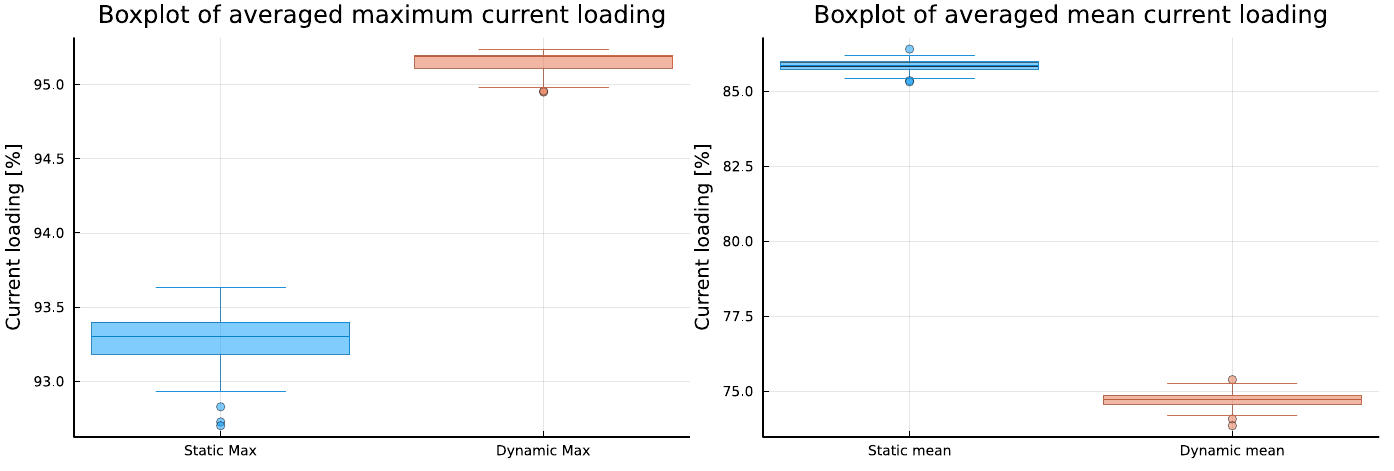}
    \caption{Distributions of line current metrics over 100 scenarios, stochastic iterative method: coordination slightly raises the maximum current but substantially lowers the mean current.}
    \label{fig:stoch_iter_i_dist}
\end{figure}

Similar to the HC distributions, the stochastic iterative approach also allows us to estimate distributions of the key power flow metrics. The left and right plots in \cref{fig:stoch_iter_v_dist} show the distributions of the maximum and mean voltage over the network, when averaged over all intermediate penetration levels (at each iteration), respectively. We see that both the mean and maximum voltages are higher in the dynamic case. This is expected since dynamic coordination allows us to attain much higher PV penetrations. The dynamic maximum voltage varies much more across scenarios than the static maximum, and the dynamic mean voltage varies somewhat more than the static mean. This may be because the extreme voltages are more sensitive to changes in constraints. \cref{fig:stoch_iter_i_dist} shows the corresponding current distributions: the maximum current loading is slightly higher in the dynamic case (about 95\% versus 93\% of ampacity) but varies less across scenarios than in the static case. The higher worst-case values are once again due to the higher PV injections enabled by dynamic coordination. However, similar to \cref{fig:metric_changes_48_75} and \cref{fig:current_metrics_det_iter}, the dynamic mean current loading is much lower (about 75\% versus 86\%), with a spread similar to that of the static case. This is likely because the flexibilities of the HPs, EVs, and BS in solving the ACOPF problem help reroute the power flows more efficiently and also minimize resistive losses. Furthermore, coordinated operation of BS and flexible loads co-located with solar can help reduce the net injections by that node into the grid. Overall, the voltages are more affected by uncertainty in the dynamic case than in the static case, as reflected by the larger spread in both the dynamic maximum and mean voltages, whereas the spread of the current loading is similar (mean) or smaller (maximum). In general, the dynamic approach improves average outcomes while widening the absolute spread of the voltage metrics, particularly the maximum voltage; relative to the higher HC it achieves, the spread of HC itself narrows slightly (\cref{tab:stoch_iter_hc_dist}).

\subsection{Reconciling the three methods}
\label{sec:reconcile}

The three methods report different PV HC values because they answer different questions at different DER levels (\cref{tab:reconcile}). The deterministic iterative method evaluates one input scenario, with storage, heat pumps and EVs at 5\% and DERs placed in proportion to nodal load. The 2-SSP evaluates five representative scenarios at higher storage (20\%), heat pump (20\%) and EV (10\%) levels, and it optimizes where DERs are placed in both its static and dynamic cases. The stochastic iterative method repeats the deterministic iterative method over 100 sampled scenarios.

All three move in the same direction: coordination raises PV HC in every case. The deterministic and stochastic iterative results agree closely (83\% against a mean of 83.6\% with coordination, and 48\% against a mean of 51.1\% without), as expected since the latter samples around the same baseline profiles. The 2-SSP gain is the largest because its dynamic case co-optimizes siting with coordinated dispatch and starts from higher storage and flexible-load levels. Its static baseline is also the lowest, because the static siting must remain feasible in all five scenarios at once and the reduced scenario set deliberately retains the most severe cases (\cref{sec:accel_2ssp}).

\begin{table}[htbp]
    \centering
    \small
    \begin{tabular}{@{}lllllccc@{}}
    \toprule
    Method & Scenarios & Storage & HP & Siting & Static PV HC & Dynamic PV HC & Gain \\
    \midrule
    DI & 1 & 5\% & 5\% & By nodal load & 48\% & 83\% & 1.7$\times$ \\
    2-SSP & 5 (of 100) & 20\% & 20\% & Optimized & 24.2\% & 98\% & 4.0$\times$ \\
    SI & 100 & 5\% & 5\% & By nodal load & 51.1\% (mean) & 83.6\% (mean) & 1.6$\times$ \\
    \bottomrule
    \end{tabular}
    \caption{PV hosting capacity from the deterministic iterative (DI), two-stage stochastic (2-SSP) and stochastic iterative (SI) methods. The methods differ in scenarios, DER levels and whether siting is optimized, so their absolute values differ, but coordination raises PV HC in all three.}
    \label{tab:reconcile}
\end{table}

\section{Discussion and broader implications}
\label{sec:discussion}

The current conversation largely takes the perspective that load growth and renewable growth pose a challenge for grid planning and operation due to the additional stress placed on infrastructure and insufficient existing grid capacity. Expanding distribution grid capacity often requires interventions like upgrading substation transformers, increasing ampacity of lines, installing additional voltage regulators and capacitor banks, etc. It is also challenging to upgrade, retrofit, or build new infrastructure in many regions (especially the United States) due to the massive capital investment required, regulatory issues related to permitting and environmental laws, as well as opposition from local communities \cite{Gorman2025}. These also cause significant delays and increase the cost of grid projects. On the other hand, dynamic coordination provides an appealing alternative by intelligently managing and optimizing resources and energy assets using virtual software-based solutions, to either replace or augment investments in new physical infrastructure. Our results indicate that with dynamic coordination and co-location, as we increase penetrations of the different DER types, their complementarity benefits can become stronger and help reduce grid stress in terms of voltage and current constraints. In addition to boosting hosting capacity, this allows DERs to capture more economic value from participating in current and future market structures.

\subsection{Value of deferred grid upgrades}
\label{sec:deferral}

A first-order estimate of the value of coordination multiplies the extra capacity it hosts by the cost of the reinforcement that capacity would otherwise require. In the deterministic iterative results, coordination adds $(83\% - 48\%) \times 2{,}930~\text{kW} \approx 1.0$~MW of PV to the feeder. For heat pumps, the gain from 9\% to 55\% of the feeder's approximately 840 homes, at the 3~kW unit rating used in this case study, gives $(55\% - 9\%) \times 840 \times 3~\text{kW} \approx 1.2$~MW of added heat pump load. Priyadarshan et al.\ estimate that accommodating residential electrification across the U.S.\ could require 312~GW of distribution reinforcement costing \$183--415 billion \cite{Priyadarshan2024DistributionElectrification}, or about \$590--1{,}330 per kW. PV-driven upgrade costs are strongly feeder-specific \cite{Horowitz2018TheStates}, so we apply the same range to both resources as an order-of-magnitude estimate. \cref{tab:deferral} gives the resulting deferral value: about \$0.6--1.4 million per feeder for the added PV and \$0.7--1.5 million for the added heat pump load. The two cases are separate studies and should not be summed.

\begin{table}[htbp]
    \centering
    \small
    \begin{tabular}{@{}lcccc@{}}
    \toprule
    Resource & Static HC & Dynamic HC & Added capacity & Deferral value per feeder \\
    \midrule
    PV & 48\% of peak load & 83\% of peak load & $\approx 1.0$~MW & \$0.6--1.4 million \\
    Heat pumps & 9\% of homes & 55\% of homes & $\approx 1.2$~MW & \$0.7--1.5 million \\
    \bottomrule
    \end{tabular}
    \caption{First-order value of the capacity added by coordination (deterministic iterative results), at \$590--1{,}330 per kW of avoided distribution reinforcement \cite{Priyadarshan2024DistributionElectrification}. The value is also the break-even budget for the coordination layer on this feeder.}
    \label{tab:deferral}
\end{table}

Implementing coordination requires sensing, monitoring, communication and DER orchestration systems \cite{Horowitz2018TheStates}. Rather than assume a cost for this layer, we report the break-even: coordination pays for itself on this feeder if the cost of the coordination layer, spread over the capacity it adds, stays below about \$590--1{,}330 per kW, i.e., below about \$0.6--1.4 million in the PV case and \$0.7--1.5 million in the heat pump case. Additional value from lower losses, lower import costs and reduced transmission dependence is not included. The estimate also ignores the time value of deferral and differences between feeders, which a utility-specific study would need to capture.

\subsection{Deployment pathway}
\label{sec:deployment}

The coordination studied here does not have to wait for new market designs; regulators are already putting parts of it in place. In California, the Public Utilities Commission's 2024 decision on Limited Generation Profiles lets distributed solar and storage interconnect without paying grid upgrade fees if they accept export limits that follow a fixed schedule, adjustable up to 24 times a year and informed by the utility's hourly integration capacity analysis \cite{pvmag2024lgp}. Dynamic hosting capacity is the real-time limit of that idea. In South Australia, SA Power Networks' flexible-exports offer lets customers export up to 10~kW per phase, varying with available network capacity, instead of a fixed 5~kW per phase, and the full 10~kW is expected to be available about 98\% of the time \cite{ena2021flexible}; such dynamic operating envelopes are the subject of a national program of trials \cite{arena_doe}. Non-wires alternatives provide the planning precedent: Con Edison's Brooklyn--Queens Demand Management program committed about \$200 million to demand management and distributed resources to defer a substation estimated to cost \$1.2 billion or more, targeting 52~MW of relief \cite{icn2016bqdm}.

Each method in this paper maps to an existing utility process. The 2-SSP supports planning-stage siting and interconnection screening, where the planner decides where new DERs can connect; the deterministic iterative method matches the hourly integration capacity analysis used to publish hosting capacity maps; and the coordinated dispatch corresponds to a DERMS that issues operating envelopes or setpoints in real time. Coordination can be delivered through a DSO-run market, a DERMS or flexible interconnection agreements, and the results here do not depend on which. Dynamic coordination of DERs thus sits alongside other data-based grid modernization strategies to optimize both transmission and distribution grids. There is also an increasing push from regulators and policymakers to encourage utilities and grid operators to adopt such technologies, primarily to reduce prices for ratepayers and move towards new electricity rate schemes like performance- or service-based regulation. This would further incentivize dynamic approaches for managing DERs. More broadly, rather than seeing this electricity demand and generation growth as a challenge for the grid, we can reframe this as an opportunity. By intelligently incentivizing, managing, and coordinating DER flexibility, and leveraging their complementary relationships, we can better accommodate this growth while also lowering costs, congestion and losses. Such an approach that simultaneously co-optimizes for generation, load, and storage can also help accelerate the planning process and DER integration, shortening the time from interconnection request to operation.

\subsection{Limitations and future work}
\label{sec:limitations}

The results have several limitations. They come from one feeder, the IEEE 123-node system, modeled as a balanced network; one representative day of input profiles with sampled perturbations; and simplified device models, including a linear first-order thermal model for heat pumps and fixed EV availability windows. Hosting capacity values on other feeders, in other seasons and in unbalanced low-voltage networks will differ. We expect the direction of the results to carry over, because the mechanisms that produce them are generic to radial feeders: midday voltage rise and reverse flow that local storage and flexible load can absorb, evening voltage drop that inverter reactive support and load shifting can relieve, and line loading that flexible demand can redistribute. The size of the gain will depend on how much storage and flexible load a feeder has and on where its binding constraints lie.

In this study, we demonstrated the proposed framework on a medium-scale distribution grid, which allowed us to solve the stochastic programming problem by combining both stages into one large single-level formulation. This approach is reasonably fast for small to medium-scale networks, particularly when accelerated with scenario reduction and warm start. A possible extension would be to apply this framework to larger networks with more scenarios, which would benefit from methods like Benders' decomposition or SDDP. Adapting these decomposition methods to nonlinear problems with integer decision variables in the 2nd stage is an active research frontier --- most studies rely on linear formulations, and stochastic MINLPs have only been tackled by a few recent works, generally considering integer variables in the 1st stage (e.g., unit commitment). Future extensions could leverage such algorithms to study larger test cases such as the IEEE 8500-node system across many more scenarios.

A further avenue for enriching the framework is to incorporate greater realism by increasing the complexity of DER models, especially for EVs and HPs. More accurate thermal dynamics models for HPs could be considered as an alternative to our linear discretized model, while EVs could use more granular availability windows. Building on the heterogeneity in device parameters across nodes already included in this work, increasing the diversity in DER characteristics would further strengthen the generalizability of the resulting HC estimates.

Another direction is to sample more accurate scenarios for the stochastic approaches by considering probability distributions specific to each DER type, complementing our approach of deriving profiles from historical data under Gaussian noise. For instance, past works have explored beta distributions for solar irradiance, Ornstein-Uhlenbeck processes for electricity prices or load profiles, and exponential inter-arrival distributions for EV charging sessions. Finally, a further opportunity is to explore the hosting capacity implications and coordination benefits of integrating small, modular data centers into distribution grids. Such edge-computing sites are becoming increasingly popular for latency-sensitive and localized applications such as 5G, autonomous driving, and other Internet of Things (IoT) use cases --- these sites also open new avenues for demand flexibility.

\section{Conclusion}
\label{sec:conclusion}

We quantified the hosting capacity gained when solar PV, battery storage, heat pumps and EVs on a distribution feeder are dispatched together by network-aware coordination, compared with static operation of the same devices. Coordination, implemented as receding-horizon optimization at 5-minute resolution, uses the flexibility of storage and electrified loads to relieve the voltage and line-loading limits that otherwise cap how much new solar generation and load a feeder can accept.

With storage, heat pumps and EVs at 5\%, coordination raised PV HC from 48\% to 83\% of peak load (73\% higher) in the deterministic iterative study; with 10\% storage, 20\% PV and 5\% EVs, it raised heat pump HC from 9\% to 55\% of homes. When DER siting and sizing were co-optimized with the 2-SSP, PV HC rose from 24\% to 98\%, and the jointly feasible region of PV, heat pump and storage penetrations grew about 22-fold. Coordination turned storage and heat pumps from substitutes into complements: with storage up to 100\% of peak load, the feeder hosted PV up to about 200\% of peak load and heat pumps in up to 90\% of homes. Storage was the main driver of PV hosting capacity, while heat pump flexibility provided most of the relief on line loading. EVs, mostly away from home during the midday hours that limit PV, contributed less in the cases studied and were not included in the sensitivity analysis. Under input uncertainty, coordination raised mean PV HC by 64\%; its absolute spread widened (standard deviation from 2.4 to 3.4 percentage points) while its spread relative to the mean narrowed slightly (from 4.6\% to 4.1\%).

At published reinforcement costs of about \$590--1{,}330 per kW, the added capacity is worth roughly \$0.6--1.4 million (PV) and \$0.7--1.5 million (heat pumps) per feeder in deferred upgrades, which also sets the break-even budget for a coordination layer. Flexible interconnection rules already in use, such as scheduled export profiles in California and dynamic operating envelopes in South Australia, offer a path to capture this value. Because coordinated hosting capacity varies more in absolute terms across scenarios, planners should size it to a low percentile of its distribution, and regulators and operators adopting dynamic interconnection will need to manage that variability.

\section*{CRediT authorship contribution statement}

\textbf{Vineet Jagadeesan Nair}: Conceptualization, Data curation, Formal analysis, Investigation, Methodology, Resources, Software, Validation, Visualization, Writing – original draft, Writing – review and editing. \textbf{Morteza Vahid Ghavidel}: Conceptualization, Methodology, Supervision, Writing – original draft, Writing – review and editing. \textbf{Anuradha M. Annaswamy}: Conceptualization, Project administration, Funding acquisition, Supervision, Writing – review and editing.

\section*{Declaration of competing interest}
The authors declare that they have no known competing financial interests or personal relationships that could have appeared to influence the work reported in this paper.

\section*{Declaration of generative AI and AI-assisted technologies in the manuscript preparation process}
During the preparation of this work, the authors used Claude (Anthropic) and Codex (ChatGPT) in order to revise and edit the manuscript text, check consistency between the text, figures and simulation outputs, and format the reference list. After using these tools, the authors reviewed and edited the content as needed and take full responsibility for the content of the published article.

\section*{Acknowledgments}
This research was supported by the MIT Energy Initiative.

\section*{Data availability}
Detailed descriptions of the model equations, variables, parameters, major input and output data, and corresponding data sources are available in the main text. Any additional data needed will be made available on request.

\appendix

\section{Implementation of DER presence at nodes in 2-SSP}
\label{app:der_presence}

As an example, suppose we want to enforce the BS state-of-charge (SOC) constraints only at nodes $i$ where BS is present, i.e., where $x_i^{BS} > 0$. We introduce a binary indicator $z_i^{BS} \in \{0,1\}$ for each node, with $z_i^{BS} = 1$ if $x_i^{BS} > 0$ and $z_i^{BS} = 0$ if $x_i^{BS} = 0$, and link it to the penetration through
\begin{gather*}
    \epsilon \, z_i^{BS} \leq x_i^{BS} \leq \overline{X} \, z_i^{BS},
\end{gather*}
where $\overline{X}$ is an upper bound on the nodal penetration ($\overline{X} = 5$, i.e., 500\% of nodal peak load, in our implementation; for EVs and heat pumps the bound $x_i \leq 1$ also applies) and $\epsilon$ is a small positive tolerance ($10^{-10}$). The right-hand inequality forces $z_i^{BS} = 1$ whenever $x_i^{BS} > 0$, and the left-hand inequality forces $x_i^{BS} \geq \epsilon$ whenever $z_i^{BS} = 1$. Constraints that should hold only where BS is present are then relaxed with big-M terms in $(1 - z_i^{BS})$. For the SOC update, written as $SOC^{BS}_i(t+1) = h_{bs}(SOC^{BS}_i(t))$,
\begin{gather*}
    SOC^{BS}_i(t+1) \geq h_{bs}(SOC^{BS}_i(t)) - M(1 - z_i^{BS}), \quad SOC^{BS}_i(t+1) \leq h_{bs}(SOC^{BS}_i(t)) + M(1 - z_i^{BS}),
\end{gather*}
which reduce to the equality when $z_i^{BS} = 1$ and are inactive when $z_i^{BS} = 0$. The SOC bounds can be written directly with the indicator, which also sets the SOC to zero at nodes without storage:
\begin{gather*}
    z^{BS}_i \underline{SOC}_i^{BS} \leq SOC^{BS}_i(t) \leq z^{BS}_i \overline{SOC}_i^{BS}.
\end{gather*}
EVs are treated in the same way with an indicator $z_i^{EV}$. For heat pumps, with an indicator $z_i^{HP}$ defined analogously, the indoor temperature bounds and dynamics are relaxed at nodes without a heat pump:
\begin{gather*}
    \underline{T}_i^{in} - M(1-z^{HP}_i) \leq T^{in}_i(t) \leq \overline{T}^{in}_i + M(1-z^{HP}_i), \\
    -M(1-z^{HP}_i) \leq T^{in}_i(t+1) - \theta_i T_i^{in}(t) - (1-\theta_i)\left(T^{out}_i(t) + \rho_i P^{HP}_i(t)\right) \leq M(1-z^{HP}_i).
\end{gather*}
At nodes without a given DER, its power variables are zero because the device ratings $\overline{P}_i^{BS}$, $\overline{P}_i^{EV}$ and $\overline{P}_i^{HP}$ are proportional to the corresponding penetrations. The constant $M$ must be large enough not to cut off feasible solutions but should be no larger than necessary, to avoid numerical difficulties; we set it from the physical range of the constrained quantity (e.g., the width of the indoor temperature band).

\bibliographystyle{elsarticle-num}
\bibliography{references, refs_manual}

\end{document}